\documentclass[a4paper,11pt]{article}
\pdfoutput=1 

\usepackage{jheppub} 

\usepackage[T1]{fontenc}
\usepackage[colorinlistoftodos]{todonotes}
\usepackage{subcaption}
\usepackage{bm}
\usepackage{afterpage}
\usepackage{sidecap}
\sidecaptionvpos{figure}{c}

\newcommand{\reffig}[1]{Fig.~\ref{#1}}
\newcommand{\refeq}[1]{Eq.~\ref{#1}}

\newcommand{\refsec}[1]{Sec.~\ref{#1}}
\newcommand{\refapp}[1]{App.~\ref{#1}}

\newcommand{\fig}[1]{\reffig{#1}}

\newcommand{\reals}{\ensuremath{\mathbb{R}}}

\newcommand{\integers}{\ensuremath{\mathbb{Z}}}

\newcommand{\lagrangian}{\ensuremath{\mathcal{L}}}
\newcommand{\manifold}{\ensuremath{\mathcal{M}}}
\newcommand{\set}{\ensuremath{\mathcal{S}}}
\newcommand{\momenta}[2]{\ensuremath{\Pi^{#1}_{\phantom{#1}{#2}}}}

\newcommand{\floor}[1]{\ensuremath{\left\lfloor #1 \right\rfloor}}
\newcommand{\avg}[1]{\ensuremath{\langle#1\rangle}}
\newcommand{\normalorder}[1]{\ensuremath{{:}\!\mathrel{#1}\!{:}}}

\newcommand{\qhat}{\hat{q}}

\title{A semi-classical recipe for wobbly limp noodles in partonic soup}

\author{R.~W.~Moerman}
\author{and W.~A.~Horowitz}
\affiliation{Department of Physics\\ University of Cape Town\\ Private Bag X3, Rondebosch 7701, South Africa}

\emailAdd{mrmrob003@myuct.ac.za}
\emailAdd{wa.horowitz@uct.ac.za}

\abstract{
We compute the average squared distance, $s^2(t)$, travelled by a light-flavour off-mass-shell coloured parton in a strongly-coupled $\mathcal{N}=4$ $SU(N_c)$ super-symmetric Yang Mills plasma using the gauge/string duality. In fact, we derive a closed integral expression for $s^2(t;a)$  in $AdS_3$-Schwarzschild, which interpolates between a heavy quark when $a = 0$ and a light quark when $a = 1$, that we evaluate analytically for small virtualities - labelled $s_\text{small}^2(t;a)$. For arbitrary virtualities, we show that for asymptotically early times the motion is ballistic, $\left.s^2(t;a)\right|_{t\ll\beta}\sim t^2$, while at asymptotically late times the motion is diffusive, $\left.s^2(t;a)\right|_{t\gg\beta}=s_\text{small}^2(t;a)\sim 2D(a) t$, from which we are able to extract the diffusion coefficient $D(a)$. Motivated by the apparent universality of the late time behaviour, we compute $s_{\text{small}}^2(t;a,d)$ and $D(a,d)$ for an arbitrary $AdS_d$-Schwarzschild geometry. From $D(a,d)$ we then compute for the first time the dynamic, time-dependent transverse momentum squared per unit path length picked up by a high momentum light quark in a strongly-coupled plasma, the transport coefficient $\hat{q}(t)$, which is critically important for phenomenology in heavy ion collisions. 
}

\begin{document} 
\maketitle
\flushbottom

\section{Introduction\label{sec:intro}}
Quantum chromodynamics (QCD) is the only theory for which we may simultaneously investigate theoretically and experimentally the non-trivial, emergent many-body dynamics of a non-Abelian quantum field theory.  After 16 years of intense work, there are still basic unanswered questions regarding the relevant dynamics of the quark-gluon plasma (QGP) produced in the heavy ion collisions at the Relativistic Heavy Ion Collider (RHIC) and Large Hadron Collider (LHC).  For example, and most important, it is still not clear whether the medium created in these collisions is strongly- or weakly-coupled.

On the one hand, recent parton cascade calculations \cite{Molnar:2001ux,Lin:2001zk,Bzdak:2014dia,Koop:2015wea,He:2015hfa,Lin:2015ucn} and pQCD-based energy loss studies \cite{Gyulassy:2001nm,Vitev:2002pf,Wang:2002ri,Majumder:2004pt,Dainese:2004te,Armesto:2005mz,Wicks:2005gt,Majumder:2007ae,Zhang:2009rn,Vitev:2009rd,Schenke:2009gb,Young:2011qx,Majumder:2011uk,Horowitz:2011gd,Buzzatti:2011vt,Horowitz:2012cf,Djordjevic:2013xoa} indicate that the medium created in heavy ion collisions is best described as a weakly-coupled gas of well-defined, slightly thermally modified QCD quasiparticles of quarks and gluons.

On the other, hydrodynamic studies \cite{Teaney:2000cw,Kolb:2001qz,Hirano:2002ds,Kolb:2003dz,Teaney:2003kp,Hirano:2005wx,Hirano:2005xf,Romatschke:2007mq,Song:2007ux,Schenke:2010rr,Shen:2011eg,Gale:2012rq,Bzdak:2013zma} imply that the low momentum modes of the medium produced in heavy ion collisions rapidly thermalize into a strongly-coupled ``soup'' of partonic matter best described by the methods of AdS/CFT; for a review, see \cite{CasalderreySolana:2011us}.  Leading order jet \cite{Morad:2014xla,Casalderrey-Solana:2014bpa} and next-to-leading order heavy flavour \cite{Horowitz:2015dta} energy loss studies based on the strong-coupling approach of AdS/CFT are also in qualitative agreement with the observed suppression of hard particles.

A further comparison of various theoretical calculations to data is thus desirable in order to try to distinguish between the two basic competing pictures for the physics of the QGP created in heavy ion collisions.  In particular, strong-coupling energy loss studies are currently limited by a lack of theoretical understanding of the fluctuations in energy loss of light and sufficiently fast moving heavy (i.e.\ effectively light) partons \cite{Gubser:2006nz,CasalderreySolana:2007qw}.  

As a first step towards a better theoretical understanding of the fluctuating energy loss of light flavors, we examine for the first time the Brownian motion of off-shell light fundamental flavours initially at rest in a strongly-coupled plasma via the AdS/CFT correspondence, building on the foundational work of the Brownian motion of an initially at rest on-shell heavy quark in de Boer et al.\ \cite{deBoer:2008gu}; additional works on the Brownian motion of on-shell heavy quarks in AdS/CFT include \cite{Atmaja:2010uu,Fischler:2012ff,Atmaja:2012jg,Sadeghi:2013lka,Banerjee:2013rca,Chakrabortty:2013kra,Sadeghi:2013jja,Fischler:2014tka}.  In a similar vein, much work has focused on the Langevin dynamics of slow-moving, on-shell heavy quarks in a strongly-coupled plasma \cite{Gubser:2006nz,Kim:2007ut,Pang:2008ru,Giecold:2009cg,Gursoy:2010aa,Giataganas:2013hwa,Giataganas:2013zaa}, in which fluctuations are imposed on a classical string solution \cite{Callan:1999ki}.

Specifically, we compute the average distance squared as a function of time moved by an off-shell, light quark initially at rest in a strongly-coupled plasma.  We find that at early times the behaviour is ballistic, $s^2(t)\sim t^2$, while at late times the behaviour is diffusive, $s^2(t) =  2 D t$, with a sub-leading sub-diffusive correction, which is characteristic of scale-less or disordered systems \cite{Havlin} and has been observed in ultra-cold atoms \cite{Sagi}. 

One of the critical transport coefficients that characterize a plasma is the transverse momentum squared per unit path length imparted from the plasma to a fast moving probe particle, known as $\hat{q}$.  In perturbative calculations, $\hat{q}$ controls the radiative energy loss of the fast moving particle and much work has centered on the phenomenological extraction of $\hat{q}$ from data (see \cite{Majumder:2010qh,Burke:2013yra} and references therein).  There has been considerable interest in the calculation of $\hat{q}$ using strong-coupling techniques, with competing results with qualitatively different numerical values and velocity dependencies from the foundational works \cite{Liu:2006ug,Gubser:2006nz,D'Eramo:2010ak} and subsequent further research \cite{Buchel:2006bv,VazquezPoritz:2006ba,Caceres:2006as,Lin:2006au,Avramis:2006ip,Armesto:2006zv,Nakano:2006js,Gao:2006uf,Argyres:2006yz,Argyres:2008eg,Edelstein:2008cp,Fadafan:2008uv,Giecold:2009cg,HoyosBadajoz:2009pv,Sadeghi:2010zp,BitaghsirFadafan:2010zh,Gursoy:2010fj,Bigazzi:2011it,Chakraborty:2011gn,Giataganas:2012zy,Chernicoff:2012gu,Zhang:2012jd,Sadeghi:2013dga,Burke:2013yra,Nam:2014sva,Li:2014hja,Li:2014dsa,Naji:2015onj,Misobuchi:2015ioa,Li:2016bbh}. 

Shockingly, we find that the late time behaviour of the Brownian motion of the endpoint of our string setup is universal.  Similar to the results of \cite{Kovtun:2004de,Herzog:2006se,Son:2009vu,CasalderreySolana:2009rm} we find the behaviour is determined completely by the physics near the black hole horizon. We are then able to extract for the first time ever the time dependence of the transport coefficient $\hat{q}$ for a light quark traversing a strongly-coupled quark-gluon plasma.  Our result has the same qualitative velocity dependence as in \cite{Liu:2006ug,D'Eramo:2010ak}, but with the proportionality constant as in \cite{Gubser:2006nz}.

This paper is organized as follows.  First, we determine the leading order classical behaviour of the AdS/CFT analogue of an off-shell light quark.  Since the isothermal coordinates necessary for the next-to-leading order analysis in the second part of the paper cannot be inverted for the whole spacetime for $d>3$, we, like de Boer et al.\ \cite{deBoer:2008gu}, restrict ourselves to an $AdS_3$-Schwarzschild geometry.  It turns out that our light flavor ``limp noodle'' solution can be thought of as the heavy flavor solution of de Boer et al.\ \cite{deBoer:2008gu} with the loose endpoint of the light flavor string travelling down the heavy flavor solution at the local speed of light.  We will find it valuable to smoothly interpolate between the on-shell heavy quark solution and that of an off-shell light quark; we parameterize the fraction of the local speed of light at which the endpoint of the string falls as $a$, with $a = 0$ giving the on-shell heavy quark solution and $a = 1$ the off-shell light quark solution.

Second, we compute semi-classically the fluctuations on top of the leading order falling string solution induced by thermal Hawking radiation emitted by the black hole; see \fig{fig:equilibratingstring}.  From this next-to-leading order analysis, we derive an integral expression for the average transverse distance squared, $s^2(t;a)$, travelled by a string whose endpoint falls with a fraction $a$ of the local speed of light.  This formal expression for $s^2(t;a)$ is analytically evaluated for small initial string lengths $\ell_0$ compared to the Hawking radius of the black hole $r_H$ (which is equivalent in the field theory to a light quark of small initial virtuality compared to the temperature of the plasma, $Q^2\ll T^2$). We also analytically evaluate $s^2(t;a)$ in the limits of asymptotically early and late times.  We find that the early time behaviour is ballistic $s^2(t;a)\sim t^2$ while the late time behaviour is diffusive $s^2(t;a)\sim t$.  Astonishingly, it is the temperature alone of the plasma---independent of the initial virtuality of the light quark or $a$---that sets the scale for ``early'' and ``late''  times.

Crucially, we find the behaviour of a small initial length string at any time is identical to the late time behaviour of a string of \emph{arbitrary} initial length with, again, the time scale set only by the temperature of the plasma, independent of the initial length of the string or $a$.  Consequently, the diffusive asymptotic behaviour of a light quark of any initial virtuality is encoded in the diffusion coefficient derived from a light quark of small virtualities.  Since one may invert the isothermal coordinates near the black hole horizon in arbitrary dimension, we compute the average transverse distance squared travelled by a short string with endpoint moving with fraction $a$ of the local speed of light in $d$ dimensions, $s^2(t;a,d)$, from which we find the diffusion coefficient $D(a,d)$.  

We then connect the transverse spatial diffusion $D(a,d)$ to the transverse momentum diffusion $\hat{q}(t)$, where the rate at which the endpoint of a falling off-mass shell light quark is found through standard numerical methods \cite{Chesler:2008wd,Chesler:2008uy}. We conclude with an outlook on future work.

\section{Leading Order (LO) Limp Noodle Dynamics\label{sec:lo}}

In order to compute the (next-to-leading order) fluctuations in the classical motion of the falling string, we must first compute the (leading order) classical motion. 
For a review of the holographic setup used throughout this paper, see, e.g., \cite{CasalderreySolana:2011us}.

Our limp noodle is initially extended along the radial direction of $AdS_d$-Schwarzs\-child with one endpoint attached to the black-brane and the other endpoint allowed to fall freely from some yet-to-be specified radial slice of $AdS_d$-Schwarzschild in the exterior region of the black-brane, in the presence of a space filling D$7$-brane \cite{Karch:2002sh}. This construction is motivated by \cite{Gubser:2008as}, wherein it is discussed that at finite temperature any fundamental string holographically dual to a coloured parton acquires its fundamental and/or anti-fundamental colour charge(s) by being attached to the black-brane; i.e.~quarks have one endpoint attached to the black-brane while gluons have both endpoints attached. Consequently, our initial condition is not only valid, but necessary. 
\par
Given that the local speed of light in the transverse directions (directions parallel to the black-brane) is zero at the horizon, the point at which the limp noodle pierces the horizon cannot move in the transverse directions \cite{Gubser:2008as}. Furthermore, the regions below and above the horizon are classically, causally disconnected \cite{Gubser:2008as}. We can, therefore, neglect any dynamics below the horizon as being indistinguishable from the thermal medium and instead consider a limp noodle with the fixed endpoint attached instead to a stretched horizon at some small yet finite distance above the horizon---an infrared (IR) cut-off which later acts as an IR regulator \cite{deBoer:2008gu}.
\afterpage{\begin{figure}
\includegraphics[width=\linewidth]{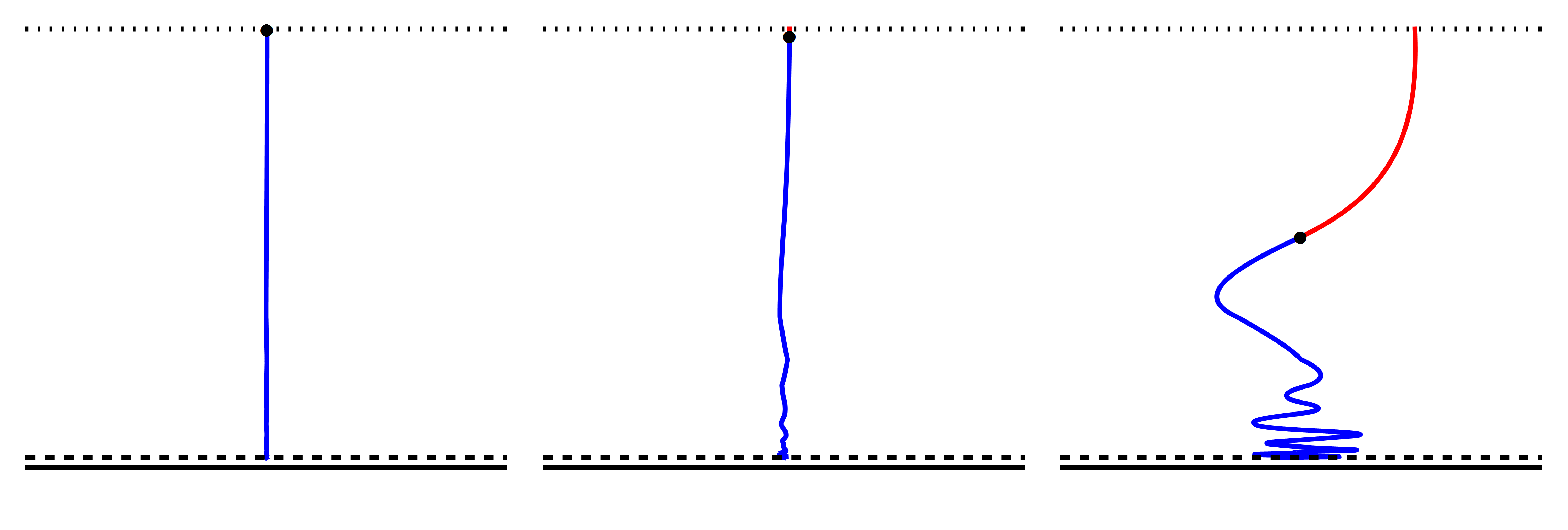}
\caption{(Colour online) Snapshots (with time increasing from left to right) of an equilibrating limp noodle (blue curve) with the free endpoint (black dot) falling from its initial position at $r=r_s+\ell_0$ (dotted black line) towards the stretched horizon at $r=r_s$ (dashed black line). The free endpoint of the limp noodle can be interpreted as an observer travelling down the stretched string of de Boer et al.\ \protect\cite{deBoer:2008gu} (red curve) at the local speed of light.}
\label{fig:equilibratingstring}
\end{figure}
}The left-most plot in \reffig{fig:equilibratingstring} shows the limp noodle (blue curve) initially extended from the stretched horizon (dashed black line) to some yet-to-be specified radial slice (dotted black line) of $AdS_d$-Schwarzschild. The middle and right-most plots shall be described later.

\subsection{Static-Gauge Polyakov String Equations of Motion\label{sec:lo-poly}}
The LO dynamics of the limp noodle are described by the Polyakov action
\begin{align}
S_\text{P} := - \frac{1}{4\pi\alpha^\prime}\int_\manifold d^2\sigma\sqrt{-\gamma}\gamma^{ab}g_{ab} = \int_\manifold d^2\sigma \lagrangian_\text{P}, \label{eq:lo-ply-action}
\end{align}
where $\gamma_{ab}$ is an auxiliary world-sheet metric, $g_{ab} := \partial_aX^\mu\partial_bX^\nu G_{\mu\nu}$ is the induced world-sheet metric and $G_{\mu\nu}$ is the space-time metric for $AdS_d$-Schwarzschild which we do not need to yet specify. We shall consider the static gauge in which we take $\sigma^a=(t,\sigma)$ to be the co-ordinates on the world-sheet parameter-space with the embedding  of the string world-sheet given by $X^\mu:[0,t_f]\times[0,\sigma_f]\to \reals^{d-1,1}$. The canonically conjugate momenta are given by 
\begin{align}
\momenta{a}{\mu} := \frac{\partial\lagrangian_\text{P}}{\partial(\partial_aX^\mu)} = -\frac{1}{2\pi\alpha^\prime}\sqrt{-\gamma}\gamma^{ab}G_{\mu\nu}\partial_bX^\nu, \label{eq:lo-ply-momentum}
\end{align}
and the energy-momentum tensor by
\begin{align}
T_{ab} := -4\pi\frac 1{\sqrt{-\gamma}}\frac{\delta S_\text{P}}{\delta \gamma^{ab}} = \frac{1}{\alpha^\prime}\left(-\frac 12\gamma_{ab}\gamma^{cd}g_{cd}+g_{ab}\right), \label{eq:lo-ply-emtensor}
\end{align}
where $\frac{\delta}{\delta \gamma^{ab}}$ is understood to be the Fr\'{e}chet derivative with respect to the auxiliary world-sheet metric. The Polyakov action has two independent dynamical variables; namely, $\gamma^{ab}$ and $X^\mu$. Requiring that the functional variation of \refeq{eq:lo-ply-action} with respect to $\gamma^{ab}$ vanishes (which is equivalent to requiring that the energy-momentum tensor vanishes) produces the constraint equation 
\begin{align}
g_{ab} = \frac 12 \gamma_{ab}\gamma^{cd}g_{cd} \iff \frac{g_{ab}}{\sqrt{-g}} = \frac{\gamma_{ab}}{\sqrt{-\gamma}}. \label{eq:lo-ply-constraint}
\end{align}
We can eliminate re-parametrisation invariance in \refeq{eq:lo-ply-action} by choosing the conformal gauge wherein we set the induced world-sheet metric to the Minkowski metric; i.e.~$\gamma^{ab}=\eta^{ab}$. With this particular choice of the auxiliary world-sheet metric, the induced world-sheet metric is conformally flat and \refeq{eq:lo-ply-constraint} reads $g_{ab} = \sqrt{-g}\eta^{ab}$, which produces the Virasoro constraint equations
\begin{align}
G_{\mu\nu}\partial_\pm X^\mu\partial_\pm X^\nu =0. \label{eq:lo-ply-constraint-conformal}
\end{align} 
The string equations of motion are obtained by requiring that the functional variation of \refeq{eq:lo-ply-action} with respect to $X^\mu$ vanishes and are given by
\begin{align}
0= \partial_a\momenta{a}{\mu} - \Gamma^{\alpha}_{\mu\nu}\partial_a X^\nu\momenta{a}{\alpha}=:\nabla_a\momenta{a}{\mu}, \label{eq:lo-ply-eom}
\end{align}
which is the string analogue of the geodesic equation for a point particle where
\begin{align*}
\Gamma^{\alpha}_{\mu\nu} := \frac 12 G^{\alpha\gamma}\left(\partial_\mu G_{\nu\gamma} + \partial_\nu G_{\mu\gamma} - \partial_\gamma G_{\mu\nu}\right),
\end{align*}
is the Christoffel symbol. The derivation of \refeq{eq:lo-ply-eom} is detailed in \refapp{app:ply}. Apart from the equations of motion, requiring that the functional variation of \refeq{eq:lo-ply-action} with respect to $X^\mu$ vanishes also imposes boundary conditions on the string endpoints,
\begin{align}
\left.\delta X^\mu\momenta{\sigma}{\mu}\right|^{\sigma = \sigma_f}_{\sigma = 0}=0.\label{eq:lo-ply-bcs}
\end{align} 
\par
At LO, the dynamics of our limp noodle are constrained to the $(1+1)$-dimensional subspace of $AdS_d$-Schwarzschild spanned by the radial and temporal directions. According to \cite{Bars:1994ek,Bars:1994sv}, the metric for any $(1+1)$-dimensional space-time (or any $(1+1)$-dimensional subspace of a higher-dimensional space-time where one of these subspace dimensions is spanned by the temporal direction) can be transformed to a metric that is conformally flat; i.e.~there exist isothermal coordinates $\{y^0,y^1\}$ such that the metric is given by
\begin{align}
ds^2 =G_{\mu\nu}dy^\mu dy^\nu=-G(y^0,y^1)(dy^0dy^1+dy^1dy^0), \label{eq:lo-bars-metric}
\end{align}
where we shall suppress the coordinate dependence of $G(y^0,y^1)$ hereafter for notational brevity.
We rewrite \refeq{eq:lo-ply-momentum} as
\begin{align}
\momenta{\pm}{\mu}&=-\frac{1}{2\pi\alpha^\prime}\eta^{\pm b}G_{\mu\nu}\partial_bX^\nu=-\frac 1{4\pi\alpha^\prime} G\partial_{\mp}Y_{1-\mu}. \label{eq:lo-bars-momentum}
\end{align}
Simplifying \refeq{eq:lo-ply-eom}
we obtain the string equations of motion and the Virasoro constraint equations for any $(1+1)$-dimensional space-time as\footnote{These are precisely the string equations of motion and constraint equations obtained in Eq.~1 of \cite{Bars:1994ek} and in Eq.~2.4 of \cite{Bars:1994sv} where $u=Y^0$ and $v=Y^1$.} 
\begin{align}
\begin{array}{rl}
0&=G\partial_+\partial_-Y^0+(\partial_{0}G)(\partial_+Y^0)(\partial_-Y^{0}), \\
0&=G\partial_+\partial_-Y^1+(\partial_{1}G)(\partial_+Y^1)(\partial_-Y^{1}), \\
0&=(\partial_\pm Y^0)(\partial_\pm Y^1),
\end{array}
\label{eq:lo-bars-eom+constraint}
\end{align}
where $\partial_{0,1}:=\frac{\partial}{\partial Y^{0,1}}$.
Using the terminology developed by Bars and Schulze \cite{Bars:1994ek,Bars:1994sv}, there exist four classes of general solutions to \refeq{eq:lo-bars-eom+constraint}. These are given by
\begin{align}
\begin{array}{rll}
\text{A:}& \quad Y^0 = Y_A^0(\sigma^+), 
& \quad Y^1=Y_A^1(\sigma^-), \\
\text{B:}& \quad Y^0 = Y_B^0(\sigma^-), 
& \quad Y^1=Y_B^1(\sigma^+), \\
\text{C:}& \quad Y^0 = Y_C^0 = \text{const}, 
& \quad Y^1=f_C(\alpha_C(\sigma^+)+\beta_C(\sigma^-); Y_C^0),  \\
\text{D:}& \quad Y^0=f_D(\alpha_D(\sigma^+)+\beta_D(\sigma^-); Y_D^0), & \quad Y^1 = Y_D^1 = \text{const}, 
\end{array} \label{eq:lo-bars-classes}
\end{align}
where $Y^0_A(\sigma^+)$, $Y^1_A(\sigma^-)$, $Y^0_B(\sigma^-)$, $Y^1_B(\sigma^+)$, $\alpha_C(\sigma^+)$, $\beta_C(\sigma^-)$,  $\alpha_D(\sigma^+)$, $\beta_D(\sigma^-)$ are arbitrary functions and $Y_C^0,Y_D^1$ are arbitrary constants. The first two classes $A,B$ exist for any metric and are easily checked by direct substitution into \refeq{eq:lo-bars-eom+constraint}. The last two classes $C,D$ are metric specific with metric specific functions $f_C,f_D$ calculated through 
\begin{align}
\begin{array}{rll}
\text{C:}& Y^0 = Y_C^0, &\quad \int^{f_C}ds G(Y^0_C,s) = \alpha_C(\sigma^+)+\beta_C(\sigma^-),  \\
\text{D:}& Y^1 = Y_D^1, &\quad \int^{f_D}ds G(s,Y^1_D) = \alpha_D(\sigma^+)+\beta_D(\sigma^-).
\end{array} \label{eq:lo-bars-functions}
\end{align}
Taking derivatives $\partial_\pm$ of \refeq{eq:lo-bars-functions} via the Liebniz integral rule yield the following relations
\begin{align*}
\begin{array}{rll}
\text{C:}& \partial_\pm Y^0 = 0, &\quad G\partial_\pm Y^1 = \partial_\pm(\alpha_C(\sigma^+)+\beta_C(\sigma^-)),  \\
\text{D:}& \partial_\pm Y^1 = 0, &\quad G\partial_\pm Y^0 = \partial_\pm(\alpha_D(\sigma^+)+\beta_D(\sigma^-)), 
\end{array}
\end{align*}
which solve \refeq{eq:lo-bars-eom+constraint}. The classes $C,D$ of general solutions describe the trajectory of a massless point particle \cite{Bars:1994sv}, because for constant $Y^0$ or $Y^1$ 
the induced world-sheet metric vanishes, indicating that the classes $C,D$ of general solutions map to a null geodesic of the target space and travels at the local speed of light. 
\par
The LO world-sheet of our limp noodle is a solution to \refeq{eq:lo-bars-eom+constraint}. For physicality, we require that any solution to \refeq{eq:lo-bars-eom+constraint} be periodic in $\sigma$ with period $4\sigma_f$ and that the global time coordinate be an increasing function of $t$ for all $\sigma\in[0,\sigma_f]$ \cite{Bars:1994ek,Bars:1994sv}; the latter condition is trivially satisfied in the static gauge. These two requirements, of periodicity and forward propagation respectively, are incompatible for any single solution given by \refeq{eq:lo-bars-classes} describing the entire world-sheet. However, these two requirements can be satisfied on patches of world-sheet that are stitched together. This procedure of stitching patches together requires a priori knowledge of the following three properties:
\begin{itemize}
\itemsep0em
\item the minimal partitioning of the world-sheet into patches,
\item the correct classes of solutions for each patch,
\item and the appropriate matching conditions for stitching these patches together.
\end{itemize}
These three properties can be extracted from the limp noodle in $\reals^{1,1}$, because the patterns of classes of solutions for patches on the world-sheet are metric independent \cite{Bars:1994ek}.

\subsection{Limp Noodle in \texorpdfstring{$\reals^{1,1}$}{R\textasciicircum\{1,1\}}\label{sec:lo-mink}}

The limp noodle in $\reals^{1,1}$ is cooked-up using a recipe inspired by \cite{Ficnar:2013wba}. Consider the following static gauge ansatz for the limp noodle in $\reals^{1,1}$ 
\begin{align}
X^\mu:[0,t_f]&\times[0,\sigma_f]\to\reals^{1,1}; \nonumber\\
(t,\sigma)&\mapsto(t,x(t,\sigma)).\label{eq:lo-mink-ansatz},
\end{align}
where we require \refeq{eq:lo-mink-ansatz} to satisfy the following boundary conditions
\begin{align}
& x(t,0)=:x_0\in\reals^+,\label{eq:lo-mink-bc01}\\
&\left.\partial_\sigma  x(t,\sigma)\right|_{\sigma=\sigma_f}=0,\label{eq:lo-mink-bc02}
\end{align} 
i.e.~a Dirichlet boundary condition at $\sigma=0$ and a Neumann boundary condition at $\sigma=\sigma_f$. We choose $t_f,\sigma_f\in\reals^+$ such that the following initial conditions are satisfied
\begin{align}
x(0,\sigma_f)-x_0&=:\ell_0\in\reals^+,\label{eq:lo-mink-ic01}\\
x(t_f,\sigma)-x_0&=0,\label{eq:lo-mink-ic02};
\end{align} 
i.e.\ the string has some initial length $\ell_0$ and the string fully collapses to a point (for the first time) in time $t_f$.  
We may use the canonical energy density from \refeq{eq:lo-ply-momentum} to connect the initial length of our stretched string to the initial virtuality of the light quark,
\begin{align}
	Q^2= \frac{\lambda\ell^2_0}{4\pi^2 L^4}.\label{eq:initial-virtuality}
\end{align}
Hence the longer the initial string, the larger the initial virtuality of the light quark.

\begin{figure}[t!]
\centering
\begin{subfigure}{0.49\textwidth}
 \caption{$F(u)$}
   \includegraphics[width=1\linewidth]{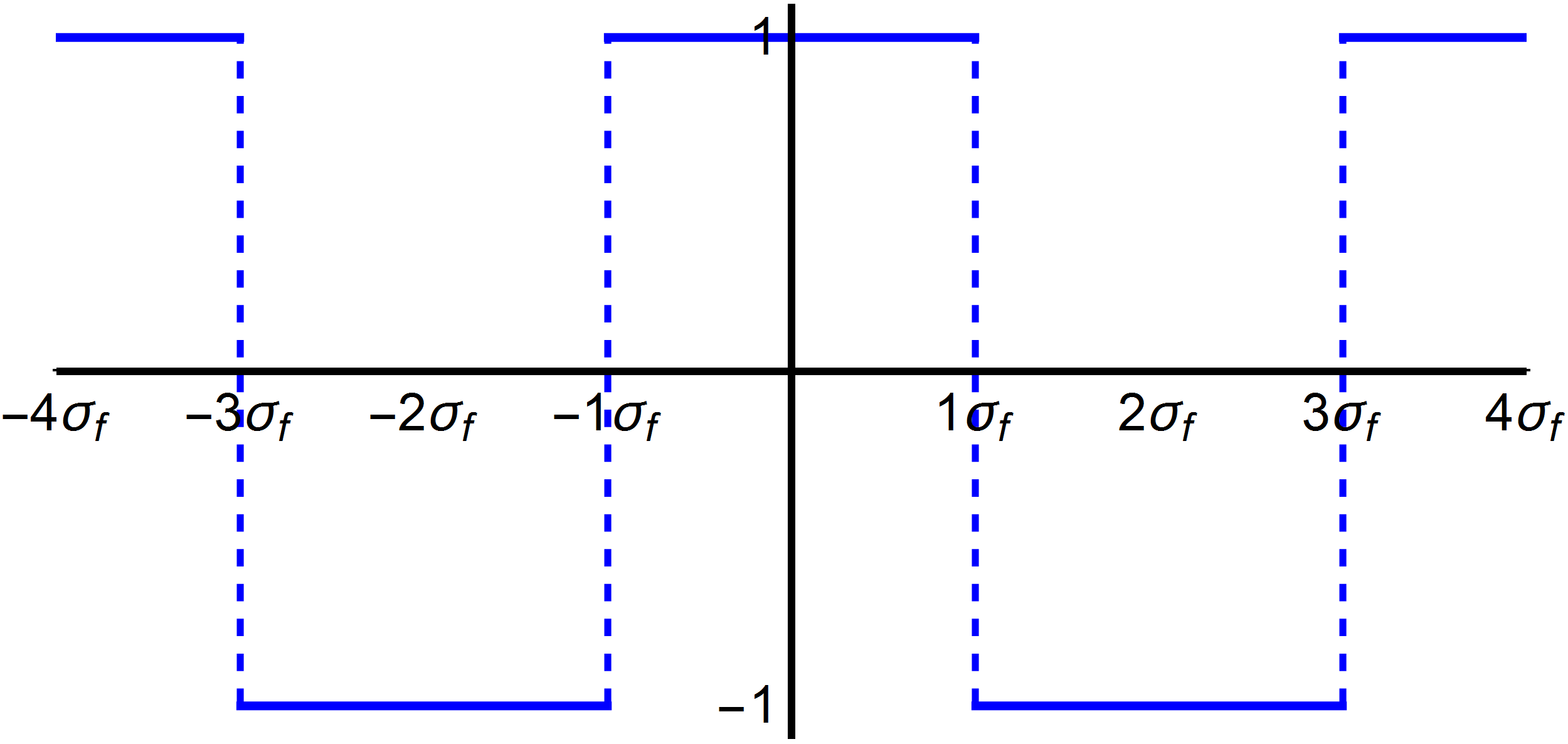}  
\end{subfigure}\hfill
\begin{subfigure}{0.49\textwidth}
   \caption{$f(u)$}
   \includegraphics[width=1\linewidth]{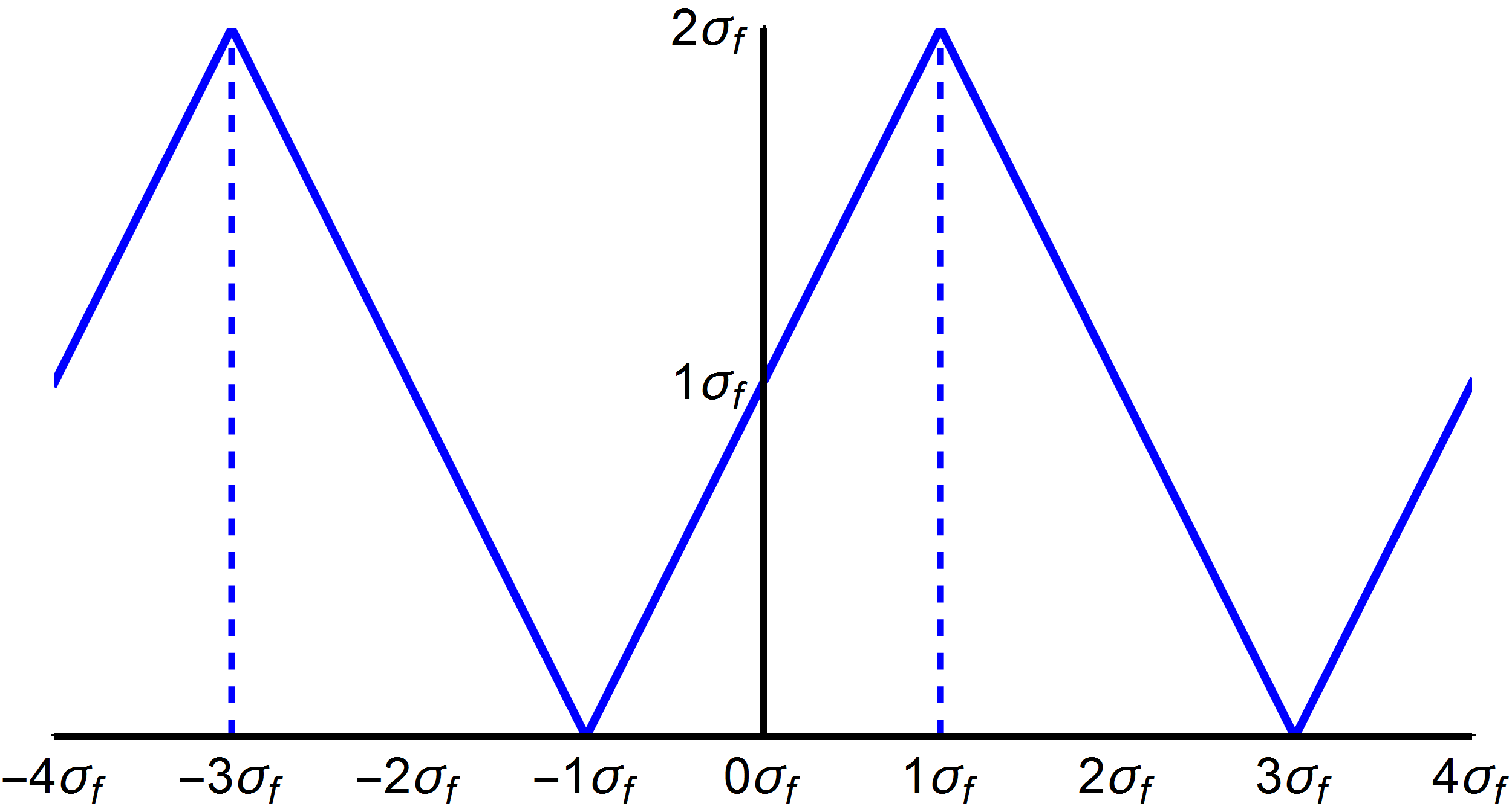}
\end{subfigure}
\caption{(Colour online) Plots of $F(u)$ (\refeq{eq:lo-mink-F}) and $f(u)$ (\refeq{eq:lo-mink-f}) over $[-4\sigma_f,4\sigma_f]$.}
\label{fig:lo-mink-buildingfuncs}
\end{figure}

\begin{figure}[t!]
\begin{subfigure}[t]{0.49\textwidth}
\centering
\includegraphics[width=0.95\textwidth]{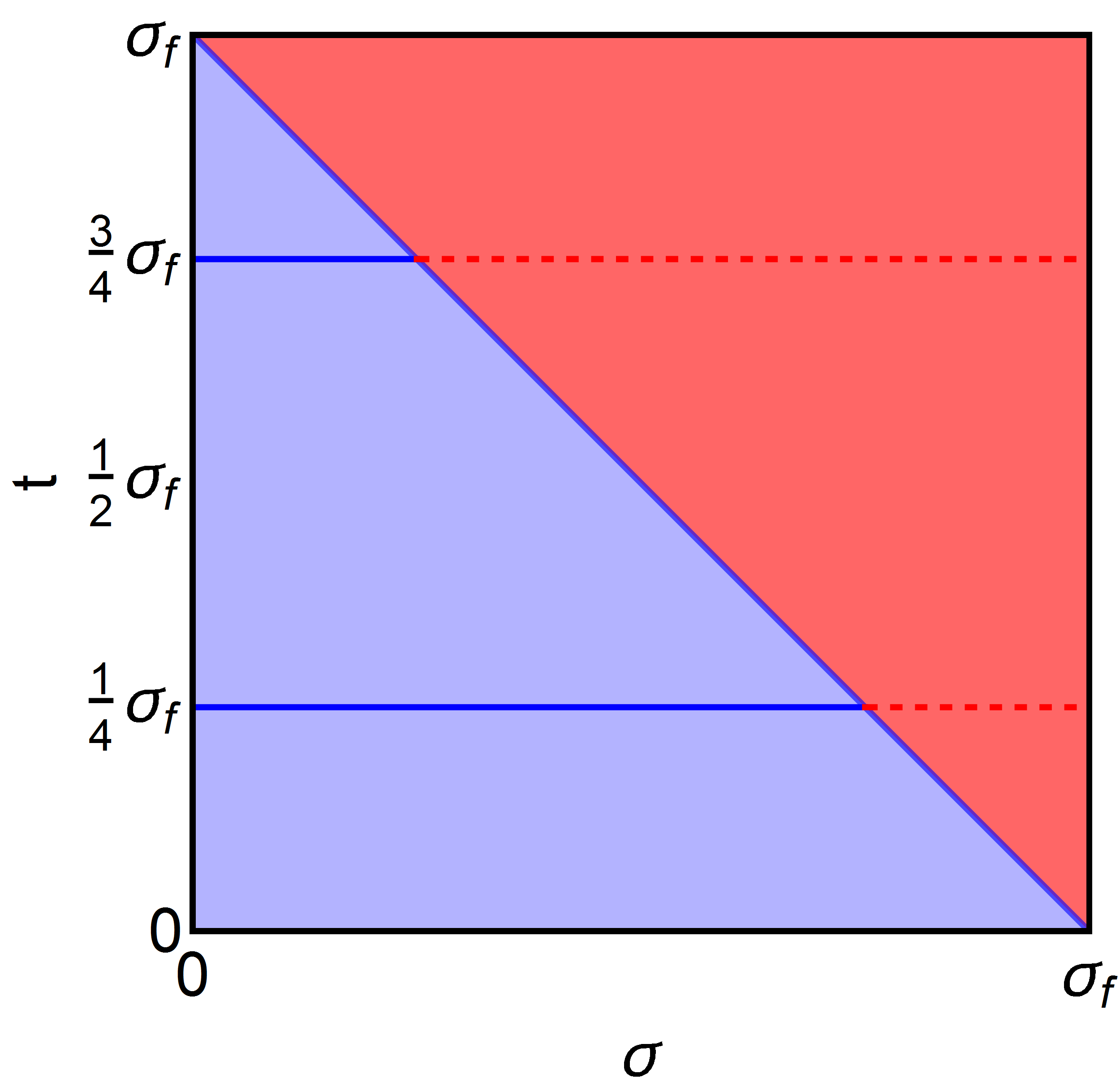}
\caption{world-sheet parameter-space $\mathcal{S}$}
\end{subfigure}
\begin{subfigure}[t]{0.49\textwidth}
\centering
\includegraphics[width=0.95\textwidth]{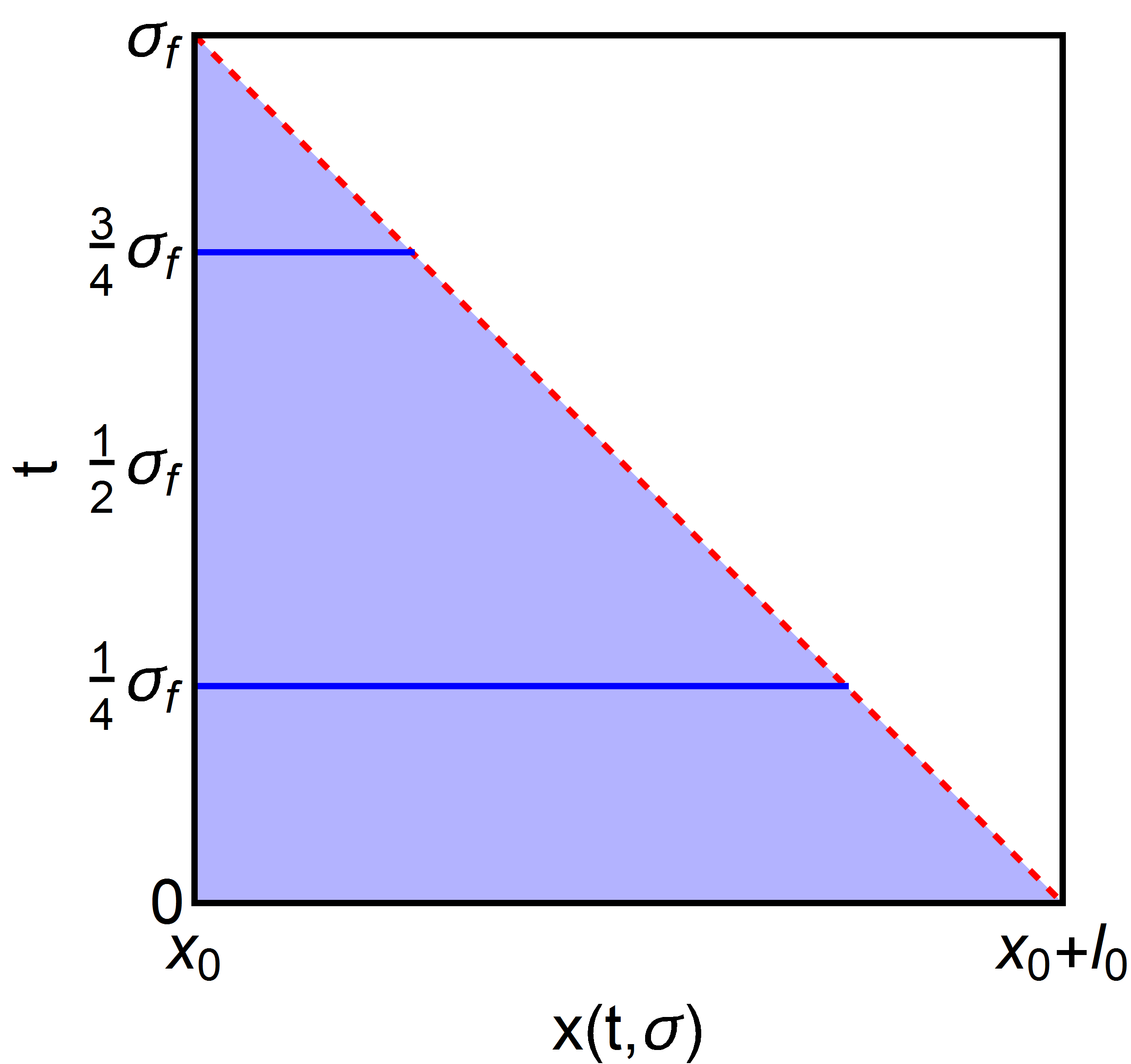}
\caption{embedding $X^\mu_\text{Mink}(\mathcal{S})$}
\end{subfigure}
\caption{(Colour online) Plots depicting $X^\mu_\text{Mink}$ (given by \refeq{eq:lo-mink-sol}) where $\set_1$ (blue shaded region) is mapped to a space-like region in $\reals^{1,1}$ (blue shaded region) and $\set_2$ (red shaded region) is mapped to a null geodesic in the target space (red dashed line).}
\label{fig:lo-mink-sol}
\end{figure}

The string equation of motion, given by \refeq{eq:lo-ply-eom}, for $x$ reads
\begin{align}
\partial_+\partial_-x(t,\sigma) = 0 \label{eq:lo-mink-eom},
\end{align}
where, for convenience, we have defined ``light-cone'' coordinates on the world-sheet parameter-space $\sigma^\pm := t\pm\sigma$. \refeq{eq:lo-mink-eom} is simply the wave equation, which has the general solution
\begin{align}
x(t,\sigma) = \frac{1}{2}(f_1(\sigma^+)+f_2(\sigma^-)), \label{eq:lo-mink-sol-gen01}
\end{align}
where $f_1,f_2:\reals\to\reals$ are continuous functions. Enforcing \refeq{eq:lo-mink-bc01} yields 
\begin{align}
x(t,\sigma) = x_0+\frac 12(f(\sigma^+)-f(\sigma^-)) \label{eq:lo-mink-sol-gen02},
\end{align}
where we have redefined $f(u):= f_1(u) = 2x_0 - f_2(u)$. Define $F(u):=\frac{d}{du}f(u)$. In terms of $F(u)$, \refeq{eq:lo-mink-bc02} becomes
\begin{align}
F(u+2 \sigma_f)=-F(u); \label{eq:lo-mink-antiperiodicity}
\end{align}
i.e.~$F(u)$ is $2\sigma_f$ anti-periodic. What remains to be satisfied are the Virasoro constraint equations, given by \refeq{eq:lo-ply-constraint-conformal}, which read
\begin{align}
F(\sigma^\pm)^2=1.\label{eq:lo-mink-constraint}
\end{align}
From \refeq{eq:lo-mink-antiperiodicity} and \refeq{eq:lo-mink-constraint} we can immediately write down 
\begin{align}
F(u) &= (-1)^{\floor{\frac{u+\sigma_f}{2\sigma_f}}}, \label{eq:lo-mink-F}\\
f(u)&=(-1)^{\floor{\frac{u+\sigma_f}{2\sigma_f}}} ((u+\sigma_f \bmod 2 \sigma_f)-\sigma_f)+\sigma_f,\label{eq:lo-mink-f}
\end{align} 
where $\bmod$ is the modulo operation and $\floor{\cdot}$ is the floor operation (see \reffig{fig:lo-mink-buildingfuncs}). 
Observe that at $t=\sigma_f$, \refeq{eq:lo-mink-sol-gen02} becomes
\begin{align*}
\nonumber x(\sigma_f,\sigma)-x_0 
&= \frac 12 (f(\sigma_f+\sigma)-f(\sigma_f-\sigma)) 
= 0.
\end{align*} 
Since we are concerned only with the behaviour of the string as it collapses from full extension, as seen in \refeq{eq:lo-mink-ic01}, to a single point at $x_0$, as seen in \refeq{eq:lo-mink-ic02}, we can choose\footnote{For the mathematical purists, we have only shown that there exists some $t_f=\sigma_f$ such that \refeq{eq:lo-mink-ic02} is satisfied, but we have not shown that $t_f=\sigma_f$ is the smallest such positive real yet. However, it is evident from \refeq{eq:lo-mink-sol} that $t_f=\sigma_f$ is indeed the smallest such positive real.} $t_f=\sigma_f$. We shall denote our square world-sheet parameter-space by $\set=[0,\sigma_f]\times[0,\sigma_f]$. Finally, our limp noodle in $\reals^{1,1}$ is given by
\begin{align}
X^\mu_\text{Mink}(t,\sigma)=\left(t, 
x_0+\left\{
\begin{array}{cl}
\sigma &, \text{if}~(t,\sigma)\in \set_1\\
\sigma_f-t &, \text{if}~(t,\sigma)\in \set_2
\end{array}
\right\}
\right), \label{eq:lo-mink-sol}
\end{align}    
where we have defined
 \begin{align}
 \begin{array}{rl}
 \set_1 &=\{(t,\sigma)\in\set|\sigma\in[0,\sigma_f-t]\}, \\
 \set_2
 &=\{(t,\sigma)\in\set|\sigma\in(\sigma_f-t,\sigma_f]\}.
 \end{array}
 \label{eq:lo-mink-ps}
 \end{align}
Notice that the world-sheet parameter-space is divided into two mutually exclusive and collectively exhaustive triangular regions, namely a lower triangular region $\set_1$ and an upper triangular region $\set_2$. This partitioning is clearly minimal (and is observed in \cite{Ficnar:2013wba}). The world-sheet embedding maps $\set_1$ to a space-like region in the target space and $\set_2$ to a null geodesic in the target space with $X^\mu_\text{Mink}(\set_2)$ on the boundary of $X^\mu_\text{Mink}(\set_1)$ as shown in \reffig{fig:lo-mink-sol}. If one transforms \refeq{eq:lo-mink-sol} to light-cone coordinates, it becomes manifestly obvious that $X^\mu_\text{Mink}|_{\set_1}$ is a solution of class $A$ while $X^\mu_\text{Mink}|_{\set_2}$ is a solution of class $C$ (see \refeq{eq:lo-bars-classes}). Remarkably, $\set_1$ maps to the entire embedded world-sheet under the action of $X^\mu_\text{Mink}|_{\set_1}$, a metric-independent solution.
\par
Suppose, we consider a modified limp noodle in $\reals^{1,1}$ in which the endpoint falls at a fraction $a$ of the local speed of light, 
\begin{align}
X^\mu_\text{Mink}(t,\sigma;a)=\left(t, 
x_0+\left\{
\begin{array}{cl}
\sigma &, \text{if}~(t,\sigma)\in \set_1(a)\\
\sigma_f-a t &, \text{if}~(t,\sigma)\in \set_2(a)
\end{array}
\right\}
\right), \label{eq:lo-mink-sola}
\end{align}
where
\begin{align}
\begin{array}{rl}
\set_1(a) &= \{ (t,\sigma)\in\set|\sigma\in[0,\sigma_f-at] \}, \\
\set_2(a)
&=\{(t,\sigma)\in\set|\sigma\in(\sigma_f-at,\sigma_f]\},
\end{array}
\label{eq:lo-mink-psa}
\end{align}
for some parameter $a\in[0,1]$. Notice first that $|\partial_t x(t,\sigma_f)|_{t>0}=a$, so $a$ is indeed the fraction of the speed of light at which the $\sigma=\sigma_f$ endpoint falls.  Notice second that $\text{area}(\set_1(a))=\left(1-\frac a2\right)\text{area}(\set)$; the fraction of the area of the parameter-space whose image is a one-to-one map onto the entire world-sheet in the target space is $1-\frac a2$. For $a=1$, \refeq{eq:lo-mink-sola} reduces to \refeq{eq:lo-mink-sol}; i.e.~$X^\mu_\text{Mink}(t,\sigma;1)=X^\mu_\text{Mink}(t,\sigma)$. For $a=0$, both \refeq{eq:lo-mink-bc02} and \refeq{eq:lo-mink-ic02} are modified to 
\begin{align}
\left.\partial_\sigma  x(t,\sigma)\right|_{\sigma=\sigma_f}&=1, \label{eq:lo-mink-bc02a}\\
x(t,\sigma)-x_0&=\sigma,\label{eq:lo-mink-ic02a}
\end{align}
where \refeq{eq:lo-mink-bc02a} tell us that we no longer have a Neumann boundary condition at $\sigma=\sigma_f$, but rather a Dirichlet boundary condition as indicated by \refeq{eq:lo-mink-ic02a}. In fact, $X^\mu_\text{Mink}(t,\sigma;0)$ we recognise as the stretched string of an at-rest, on-shell heavy quark whose mass is given by \refeq{eq:initial-virtuality}, with $Q^2$ replaced by the mass of the quark squared, $M_Q^2$. 

Introducing $a$, \refeq{eq:lo-mink-sola}, enables us to reinterpret our limp noodle in terms of the static stretched string: the free endpoint of the limp noodle analog of a static off-shell light quark corresponds to an observer travelling down the static stretched string analog of a static heavy quark at the local speed of light.

\subsection{Limp Noodle in \texorpdfstring{$AdS_3$}{AdS\_3}-Schwarzschild\label{sec:lo-ads}}
The general $AdS_d$-Schwarzschild metric for $d\ge 3$ is given by 
\begin{align}
ds_d^2 = \frac{r^2}{L^2}\left(-h(r;d)dt^2+d\bm{x}_{d-2}^2\right)+\frac{L^2}{r^2}\frac{dr^2}{h(r;d)}, \label{eq:gen-metric}
\end{align}
where $L\in\reals^+$ is the radius of curvature of $AdS_d$,
\begin{align}
h(r;d):= 1-\left(\frac{r_H}{r}\right)^{d-1},\label{eq:gen-h}
\end{align}
is the so-called blackening factor, $r_H\in\reals^+$ is the radial position of the black-brane horizon, $r\in[0,\infty)$ denotes the radial direction and $\bm{x}_{d-2}\in\reals^{d-2}$ denotes the transverse
directions parallel to the black-brane that are identified with the spatial directions in the CFT on the boundary. The Hawking temperature\footnote{For physically relevant temperatures $\sim350$ MeV observed in the fireball wake of ultra-relativistically collided heavy nuclei at RHIC and LHC, we take $r_H\sim\frac{22}{d-1}~fm$.} of the black-brane is given by
\begin{align}
T=\frac 1\beta = \frac{(d-1)r_H}{4\pi L^2}, \label{eq:gen-hawkingtemp}
\end{align}
in natural units. We define the so called tortoise coordinate \cite{carroll2004spacetime,deBoer:2008gu} by 
\begin{align}
r_\ast(d) := L^2\int\frac{dr}{r^2 h(r;d)} = -\frac{L^2}{r}{}_2F_1\left(1,\frac{1}{d-1};\frac{d}{d-1};\left(\frac{r_H}{r}\right)^{d-1}\right), \label{eq:gen-tortoisecoord}
\end{align}
where ${}_2F_1$ is the Gaussian hypergeometric function. As forwarned, \refeq{eq:gen-tortoisecoord} cannot be inverted for $d>3$. For $d=3$ we have 
\begin{align}
\begin{split}
r_\ast&:=\frac{L^2}{r_H}\coth^{-1}\left(-\frac{r}{r_H}\right),\\
r &:= -r_H \coth\left(\frac{r_H r_\ast}{L^2}\right),
\end{split} \label{eq:lo-ads-tortoisecoord}
\end{align}
and \refeq{eq:gen-metric} reduces to the $AdS_3$-Schwarzschild\footnote{The $AdS_3$-Schwarzschild metric is also the metric for the non-rotating BTZ black hole, which in standard co-ordinates has $x=L\phi$, where we make the identification $(\forall~k\in\integers)$, $\phi\simeq\phi+2\pi k$ (i.e.~$\phi\in\reals/\integers$) \cite{deBoer:2008gu}.} metric, given by 
\begin{align}
ds^2 &= -\frac{r^2-r_H^2}{L^2}dt^2 + \frac{L^2}{r^2-r_H^2}dr^2 +\frac{r^2}{L^2}dx^2, \label{eq:lo-ads-metric}\\
&= \frac{r_H^2}{L^2}\text{csch}^2\left(\frac{r_Hr_\ast}{L^2}\right)(-dt^2+dr_\ast^2)+\frac{r_H^2}{L^2}\coth^2\left(\frac{r_Hr_\ast}{L^2}\right)dx^2. \label{eq:lo-ads-metricflat}
\end{align}
We recognise the $(1+1)$-dimensional subspace of $AdS_3$-Schwarzschild spanned by the time co-ordinate and the tortoise coordinate, given by \refeq{eq:lo-ads-metricflat}, as being conformally flat. Let us introduce a stretched horizon $r_s=(1+\varepsilon)r_H$ for some $0<\varepsilon\ll 1$ just above the horizon, which acts as an IR cut-off that we shall use to regulate our calculation in \refsec{sec:nlo}. We require that the limp noodle in $AdS_3$-Schwarzschild satisfies the following boundary conditions
\begin{figure*}
\begin{subfigure}[t]{0.49\textwidth}
\centering
\includegraphics[width=0.95\linewidth]{lo-mink-ps.png}
\captionsetup{width=\textwidth} 
\caption{world-sheet $\mathcal{S}$}
\label{fig:lo-ads-sola}
\end{subfigure}\hfill
\begin{subfigure}[t]{0.49\textwidth}
\centering
\includegraphics[width=0.95\linewidth]{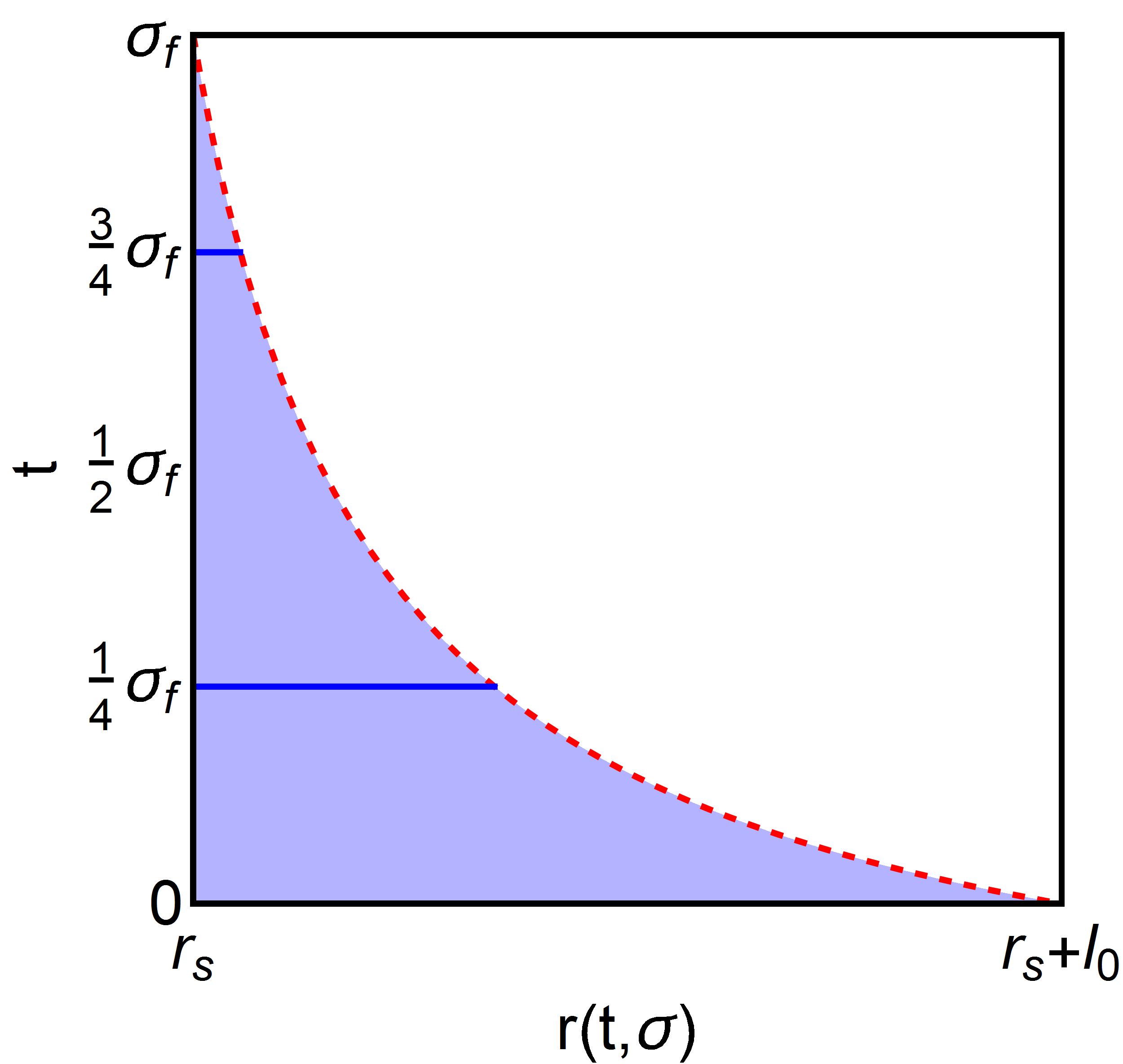}
\captionsetup{width=\textwidth} 
\caption{embedding $X^\mu_{AdS_3}(\mathcal{S})$}
\label{fig:lo-ads-solb}
\end{subfigure}
\caption{(Colour online) Plots depicting $X^\mu_{AdS_3}$ (given by \refeq{eq:lo-ads-sol}) $\set_1(1)$ (blue shaded region) mapped to a space-like region in $\reals^{1,1}$ (blue shaded region) and $\set_2(1)$ (red shaded region) mapped to a null geodesic in the target space (red dashed line) whose exact mapping (i.e.~$X^\mu_{AdS_3}|_{\set_2(1)}$) we are uninterested in.}
\label{fig:lo-ads-sol}
\end{figure*}
\begin{align}
X^r_{AdS_3}(t,0;a) &= r_s,\label{eq:lo-ads-bc01}\\
\left.\partial_\sigma  X^r_{AdS_3}(t,\sigma;a)\right|_{\sigma=\sigma_f}&=\delta_{a,0},\label{eq:lo-ads-bc02}
\end{align} 
where $\delta_{a,0}$ denotes the Kronecker delta, with $\sigma_f\in\reals^+$ chosen such that the following initial condition is satisfied
\begin{align}
X^r_{AdS_3}(0,\sigma_f;a)-r_s&=:\ell_0\in\reals^+,\label{eq:lo-ads-ic01}
\end{align} 
for some initial string length $\ell_0\in\reals^{+}$. Notice that in the conformally flat description 
of $AdS_3$-Schwarzschild, using \refeq{eq:lo-ads-tortoisecoord} we can rewrite \refeq{eq:lo-ads-bc02} in terms of the tortoise coordinate as
\begin{align*}
\left.\partial_\sigma  X^r_{AdS_3}(t,\sigma;a)\right|_{\sigma=\sigma_f} = \frac{r_H^2}{L^2} \text{csch}^2\left(\frac{r_H}{L^2}X^{r_\ast}_{AdS_3}(t,\sigma;a)\right)\left.\partial_\sigma  X^{r_\ast}_{AdS_3}(t,\sigma;a)\right|_{\sigma=\sigma_f}&=\delta_{a,0},\\
\implies \left.\partial_\sigma  X^{r_\ast}_{AdS_3}(t,\sigma;a\ne0)\right|_{\sigma=\sigma_f}&=0,
\end{align*}

since csch is never zero. Consequently, in the conformally flat description of $AdS_3$-Schwarzs-child, the boundary conditions \refeq{eq:lo-ads-bc01} and \refeq{eq:lo-ads-bc02} for $a\ne0$ correspond directly to \refeq{eq:lo-mink-bc01} and \refeq{eq:lo-mink-bc02}; for $a=0$ we produce a Dirichlet boundary condition at $\sigma =\sigma_f$. Since the appropriate boundary conditions are met, we can immediately write down the solution for the limp noodle on $\set_1(a)$ for $a\ne 0$ by transforming \refeq{eq:lo-mink-sola} with the tortoise coordinate as
\begin{align}
X^\mu_{AdS_3}|_{\set_1(a)}(t,\sigma;a) = \left(t,-r_H\coth\left(\frac{r_H({r_s}_\ast+\sigma)}{L^2}\right),0\right)^\mu, \label{eq:lo-ads-sol}
\end{align}
(see\footnote{As we decrease $a$ from $a=1$ to $0$, the topmost end of the line demarcating $\set_1(1)$ and $\set_2(1)$ of \reffig{fig:lo-ads-sola} moves from the left to the right corner. Similarly for \reffig{fig:lo-ads-solb}, the topmost end of the dashed red line moves from the left to the right corner. } \reffig{fig:lo-ads-sol} for $a=1$) 
\begin{align}
{r_s}_\ast &=\frac{L^2}{r_H}\text{coth}^{-1}\left(-\frac{r_s}{r_H}\right), \label{eq:lo-ads-r0star}\\
\sigma_f &=\frac{L^2}{r_H}\text{coth}^{-1}\left(-\frac{r_s+\ell_0}{r_H}\right)-{r_s}_\ast. \label{eq:lo-ads-sigmaf}
\end{align}
For $a=0$, $\set_1(a=0) = \set$ and \refeq{eq:lo-ads-sol} is also the solution. It turns out that knowing $X^\mu_{AdS_3}|_{\set_1(a)}$ is sufficient and we do not need to know $X^\mu_{AdS_3}|_{\set_2(a)}$ other than it maps onto $X^\mu_{AdS_3}|_{\set_1(a)}(t,\sigma=\sigma_f-at;a)$. Comparing \refeq{eq:lo-ads-sol} with the stretched string, we notice that with the identification  
\begin{align}
r=-r_H\coth\left(\frac{r_H({r_s}_\ast+\sigma)}{L^2}\right), \label{eq:lo-ads-ident}
\end{align}
the stretched string of de Boer et al.~\cite{deBoer:2008gu,Atmaja:2010uu}, which is given by 
\begin{align*}
X^\mu_{\text{dB}}(t,r)=(t,r,0)^\mu,
\end{align*}
is mapped onto \refeq{eq:lo-ads-sol}.

\section{Next-to-Leading Order (NLO) Limp Noodle Dynamics\label{sec:nlo}}
We want to gain access to the motion of a light quark induced by the thermal noise of the medium in which it resides.  According to the AdS/CFT dictionary, then, we want to understand quantitatively the motion of the string transverse to the radial direction.  

Following the semi-classical treatement of \cite{deBoer:2008gu}, we compute this transverse motion by quantizing the transverse fluctuations of our LO limp noodle and imposing a thermal average on the mode expansion.  The middle and right-most plots in \reffig{fig:equilibratingstring} depict two snapshots in the time evolution of an archetypical set of transverse fluctuations on top of the classical solution. To produce these plots we numerically evolved a discretized set of Bose-Einstein distributed fluctuations on the LO limp noodle solution at a Hawking temperature of $\sim 350$ MeV.

More generally, the dynamics of transverse fluctuations can be understood from the Nambu-Goto action
\begin{align}
S_\text{NG}:=-\frac{1}{2\pi\alpha^\prime}\int\limits_{\manifold}d^2\sigma\sqrt{-g}=\int\limits_{\manifold}d^2\sigma\lagrangian_\text{NG}, \label{eq:nlo-ng-action}
\end{align}
where $g_{ab} := \partial_aX^\mu\partial_bX^\nu G_{\mu\nu}$ is again the induced world-sheet metric and $G_{\mu\nu}$ is the space-time metric for some $d$-dimensional space-time. Still working in the static gauge, suppose we have a particular LO string solution $X^\mu_0:[0,t_f]\times[0,\sigma_f]\to\reals^{d-1,1}$ which has non-zero components in a $(1+1)$-dimensional subspace of the the space-time where one of these subspace dimensions is spanned by the temporal direction $t=x^0$ and the other subspace dimension is spanned by a spatial direction which we can choose, without loss of generality, to be $r=x^1$. We shall denote all other (transverse) directions by $x^I$ where $I=2,\cdots,d-1$, and we shall assume that $G_{\mu\nu}$ is independent of these transverse directions.
\par
Suppose that we now want to add transverse fluctuations $X^I$ (for $I=2,\cdots,d-1$) to our particular static gauge LO string solution $X^\mu_0$. Expanding \refeq{eq:nlo-ng-action} about $X^\mu_0$ in terms of these $X^I$ we obtain
\begin{align}
S_\text{NG} &= \int\limits_{\manifold}d^2\sigma\lagrangian_\text{NG}|_{X^\mu_0} + S_\text{NG}^{(2)} + \mathcal{O}\left((X^I)^3\right), \label{eq:nlo-ng-expansion}
\end{align}
where the effective action for the transverse fluctuations is given by 
\begin{align}
\nonumber S_\text{NG}^{(2)}:&= \frac 12 \int\limits_{\manifold}d^2\sigma \left.\frac{\partial^2 \lagrangian_\text{NG}}{\partial(\partial_aX^I)\partial(\partial_bX^J)}\right|_{X^\mu_0}\partial_aX^I\partial_bX^J \\
 &= -\frac{1}{4\pi\alpha^\prime}\int\limits_{\manifold}d^2\sigma\left.\left[\sqrt{-g}g^{ab}G_{IJ}\right]\right|_{X^\mu_0}\partial_aX^I\partial_bX^J = \int\limits_{\manifold}d^2\sigma\lagrangian_\text{NG}^{(2)}, \label{eq:nlo-ng-action2}
\end{align} 
which is nothing but an effective Polyakov action\footnote{\refeq{eq:nlo-ng-action2} can be thought of as describing a field theory of free massless scalars living on the world-sheet of the classical L0 string solution $X^\mu_0$.} for the transverse fluctuations where the auxiliary world-sheet metric is chosen to be the induced world-sheet metric evaluated on the LO string solution  $X^\mu_0$. This quadratic approximation for the transverse fluctuations is valid provided the fluctuations remain sufficiently small and the string endpoint remains sufficiently above the black hole horizon\footnote{For an $AdS_d$-Schwarzschild geometry, $d\ge 3$, the quadratic approximation breaks down within $\sqrt{\alpha^\prime}=L\lambda^{-1/4}$ of the black hole horizon \cite{deBoer:2008gu}.}. The canonically conjugate momenta are given by
\begin{align}
\momenta{a}{I} = \frac{\partial\lagrangian_\text{NG}^{(2)}}{\partial(\partial_aX^I)} = -\frac{1}{2\pi\alpha^\prime}\left.\left[\sqrt{-g}g^{ab}G_{IJ}\right]\right|_{X^\mu_0}\partial_bX^J. \label{eq:nlo-ng-momenta}
\end{align}
The equations of motion for the transverse fluctuations are obtained by requiring that the functional variation of \refeq{eq:nlo-ng-action2} with respect to $X^I$ vanishes and are given by
\begin{align}
0=\partial_b\left(\left.\left[\sqrt{-g}g^{ab}G_{IJ}\right]\right|_{X^\mu_0}\partial_aX^I\right) \label{eq:nlo-ng-eom},
\end{align}
subject to the boundary conditions
\begin{align}
\left.\left.\left[\sqrt{-g}g^{\sigma b}G_{IJ}\right]\right|_{X^\mu_0}\partial_aX^I\delta X^J\right|_{\sigma=0}^{\sigma=\sigma_f}=0. \label{eq:nlo-ng-bcs}
\end{align}
The derivation of \refeq{eq:nlo-ng-eom} is performed more explicitly in \refapp{app:ng}.
\par
Returning to our limp noodle in $AdS_3$-Schwarzschild, we relabel \refeq{eq:lo-ads-sol} as $X^\mu_{0}$ and add non-zero transverse fluctuations in the $x$-direction as follows
\begin{align*}
X^\mu_{AdS_3}|_{\set_1(a)}(t,\sigma;a) &= X^\mu_{0}(t,\sigma;a)+\delta^{\mu 2}X(t,\sigma) \\
&= \left(t,-r_H\coth\Big(\frac{r_H({r_s}_\ast+\sigma)}{L^2}\Big),X(t,\sigma)\right)^\mu.
\end{align*}
For our particular LO limp noodle solution, \refeq{eq:nlo-ng-eom} simplifies to
\begin{align}
0 &= -\partial_t^2X(t,\sigma) + \frac{1}{\coth^2\left(\frac{r_H}{L^2}({r_s}_\ast+\sigma)\right)}\partial_\sigma\left(\coth^2\left(\frac{r_H}{L^2}({r_s}_\ast+\sigma)\right)\partial_\sigma X(t,\sigma)\right) \label{eq:nlo-eomsig} \\
&= -\partial_t^2 X(t,r) + \frac{r^2-r_H^2}{L^4r^2}\partial_r\left(r^2(r^2-r_H^2)\partial_r X(t,r)\right), \label{eq:nlo-eom}
\end{align}
where in the last line we used \refeq{eq:lo-ads-ident}. However, \refeq{eq:nlo-eom} is precisely Eq.~2.32 of \cite{deBoer:2008gu} and is solved therein. We shall briefly discuss the salient features of the solution to \refeq{eq:nlo-eom} in the following paragraph. For notational consistency with \cite{deBoer:2008gu}, we shall use the standard coordinate description for $AdS_3$-Schwarzschild given by \refeq{eq:lo-ads-metric} in the following paragraph. However, to reduce notational clutter, we shall use function arguments as seen in  $X(t,r)$ and $X(t,\sigma)$ to denote the relevant coordinate description.
\par
\refeq{eq:nlo-eom} is linear and homogeneous. Consequently, \refeq{eq:nlo-eom} is solved by the separable ansatz
\begin{align}
X(t,r) = f_\omega(r)e^{-i\omega t},\label{eq:nlo-ansatz01}
\end{align}
for some frequency $\omega\in\reals$, and 
\begin{align}
0=\omega^2f_\omega(r)+ \frac{r^2-r_H^2}{L^4r^2}\partial_r\left(r^2(r^2-r_H^2)\partial_r f_\omega(r)\right). \label{eq:nlo-eom-ansatz}
\end{align}
Being a second order ODE, \refeq{eq:nlo-eom-ansatz} has two linearly independent solutions
\begin{align}
f^{(\pm)}_\omega(r) = \frac{1}{1\pm i\nu}\frac{r\pm i r_H\nu}{r}\left(\frac{r-r_H}{r+r_H}\right)^{\pm i \nu/2} = \frac{1}{1\pm i\nu}\frac{r\pm i r_H\nu}{r}e^{\pm i \omega (r_{s\ast}+\sigma)},\label{eq:nlo-pm}
\end{align}
where we have defined the dimensionless quantity
\begin{align}
\nu:=\frac{L^2\omega}{r_H}. \label{eq:nlo-dimless-nu}
\end{align}
In the near-horizon limit (i.e.~$r\to r_H$), $f^{(\pm)}_\omega$ are normalised such that the so-called infalling and out-going horizon boundary conditions are satisfied 
\begin{align}
f_\omega^{(\pm)}(r)\xrightarrow{r\to r_H}e^{\pm i \omega(r_{s\ast}+\sigma)}, \label{eq:nlo-bcs}
\end{align}
where the superscript $(\pm)$ naturally labels out-going (progressive) and infalling (regressive) modes respectively. Rewriting $f_\omega$ as a linear combination of the linearly independent modes we obtain
\begin{align}
f_\omega(r) = f_\omega^{(+)}(r)+B_\omega f_\omega^{(-)}(r), \label{eq:nlo-ansatz02}
\end{align}
where $B_\omega$ measures the phase shift between the infalling and out-going modes. Enforcing a Neumann boundary condition in the direction transverse to the black hole at the free endpoint\footnote{The Neumann boundary condition is enforced at $\sigma=\sigma_f$ as for the static string solution of de Boer et al.~because we were able to reinterpret (at leading order) the free endpoint of our falling string solution as an observer moving down their static string solution at the local speed of light.} fixes $B_\omega$
\begin{align}
\partial_r f_\omega(r)|_{r=r_s+\ell_0}=0\implies B_\omega = 
\frac{1-i\nu}{1+i\nu}\frac{1+i \tilde{r}_0\nu}{1-i \tilde{r}_0\nu}\left(\frac{ \tilde{r}_0-1}{ \tilde{r}_0+1}\right)^{i\nu}, \label{eq:nlo-B}
\end{align}
where we have defined the dimensionless quantity
\begin{align}
 \tilde{r}_0:=\frac{r_s+\ell_0}{r_H}. \label{eq:nlo-dimless-rho0}
\end{align}
The general solution is then constructed as a superposition of $f_\omega$ over all non-negative frequencies $\omega\in\reals$
\begin{align}
X(t,r) = \int\limits_{0}^{\infty}\frac{d\omega}{2\pi}A_\omega\left[f_\omega(r)e^{-i\omega t}a_\omega+f_\omega^\ast(r)e^{i\omega t}a_\omega^\ast\right], \label{eq:nlo-gensol-c}
\end{align}
where $a_\omega,a_\omega^\ast$ are Fourier coefficients, and $A_\omega$ is an as yet-to-be-determined normalisation constant.
According to \cite{Atmaja:2010uu}, the normalisation constant is determined by the near-horizon region where the tortoise coordinate, given by \refeq{eq:lo-ads-tortoisecoord}, becomes semi-infinite. Consequently, the normalisation is universal and calculated by requiring consistent Dirac quantisation of the operator expressions for $X$ and its canonically conjugate momentum (see \refeq{eq:nlo-ng-momenta}) which are given by 
\begin{align}
\hat{X}(t,\sigma) :=& \int\limits_{0}^{\infty}\frac{d\omega}{2\pi}A_\omega\left[f_\omega(\sigma)e^{-i\omega t}\hat{a}_\omega+f_\omega^\ast(\sigma)e^{i\omega t}\hat{a}_\omega^\dagger\right], \label{eq:nlo-norm-q} \\
\hat{\mathcal{P}}^t(t,\sigma) :=& -\frac{i}{2\pi\alpha^\prime}\frac{r_H^2}{L^2} \coth^2\left(\frac{r_H}{L^2}(r_{s\ast}+\sigma)\right) \int\limits_{0}^{\infty}\frac{d\omega}{2\pi}\omega A_\omega\left[f_\omega(\sigma)e^{-i\omega t}\hat{a}_\omega-f_\omega^\ast(\sigma)e^{i\omega t}\hat{a}_\omega^\dagger\right]
.\label{eq:nlo-norm-p}
\end{align}
Note that in \refeq{eq:nlo-norm-q} and \refeq{eq:nlo-norm-p} we have returned to the conformally flat description of $AdS_3$-Schwarzschild given by \refeq{eq:lo-ads-metricflat}.
Firstly, we enforce canonical commutation relations on the above defined operator expressions themselves
\begin{align}
\begin{split}
&[\hat{X}(t,\sigma),n_t\hat{\mathcal{P}}^t(t,\sigma^\prime)]_{\Sigma} = i\delta(\sigma,\sigma^\prime)= i\frac{\delta(\sigma-\sigma^\prime)}{\sqrt{g|_\Sigma}}, \\
&[\hat{X}(t,\sigma),\hat{X}(t,\sigma^\prime)]_{\Sigma}=0=[n_t\hat{\mathcal{P}}^t(t,\sigma),n_t\hat{\mathcal{P}}^t(t,\sigma^\prime)]_{\Sigma},
\end{split}\label{eq:nlo-norm-comrels01}
\end{align}
where $\delta(\sigma-\sigma^\prime)$ denotes the Dirac delta function, $\Sigma$ is a Cauchy surface within the world-sheet of the LO solution $X^\mu_0$ (given by \refeq{eq:lo-ads-sol}), which is chosen for convenience to be a constant time Cauchy surface, $g|_\Sigma = g_{r_\ast r_\ast}$ is the induced metric on $\Sigma$ and $n_\mu=\delta_{\mu t}/\sqrt{-g_{tt}} = n_\mu=\delta_{\mu t}/\sqrt{g_{r_\ast r_\ast}} = \delta_{\mu t}/\sqrt{g|_\Sigma}$ is the future pointing unit normal to $\Sigma$ \cite{deBoer:2008gu,Atmaja:2010uu}.The normalisation $A_\omega$ is then fixed by additionally enforcing canonical ``creation'' and ``annihilation'' commutation relations on $\hat{a}_\omega,\hat{a}_\omega^\dagger$ \begin{align}
\begin{split}
&[\hat{a}_\omega,\hat{a}_{\omega^\prime}^\dagger]_\Sigma = 2\pi\delta(\omega-{\omega^\prime}),\\
&[\hat{a}_\omega,\hat{a}_{\omega^\prime}]_\Sigma =0=
[\hat{a}_\omega^\dagger,\hat{a}_{\omega^\prime}^\dagger]_\Sigma,
\end{split}\label{eq:nlo-norm-comrels02}
\end{align}
and requiring consistency between \refeq{eq:nlo-norm-comrels01} and \refeq{eq:nlo-norm-comrels02}. Using \refeq{eq:nlo-norm-q} and \refeq{eq:nlo-norm-p} we can write
\begin{align}
[\hat{X}(t,\sigma),n_t\hat{\mathcal{P}}^t(t,\sigma^\prime)]=
\frac{1}{\sqrt{g|_\Sigma}}\frac{i}{\pi\alpha^\prime}\frac{r_H^2}{L^2}\int\limits_{-\infty}^\infty\frac{d\omega}{2\pi}A_{\omega}^2\omega\left[e^{i\omega(\sigma-\sigma^\prime)} + e^{i\omega(\sigma+\sigma^\prime)}\right], \label{eq:nlo-norm-comrels03}
\end{align}
where we have taken the limits $\sigma\to0,~r_{s\ast}\to-\infty$. For later use, we recall the definition for the $AdS$ radius of curvature $L$ in terms of the 't Hooft coupling $\lambda$ 
\begin{align}
L^4=\lambda{\alpha^\prime}^2 \label{eq:nlo-lambda}
\end{align}
If we make the ansatz that 
\begin{align}
A_\omega := \frac{L}{r_H}\sqrt{\frac{\pi\alpha^\prime }{\omega}}=\frac{\beta}{2\sqrt{\pi\omega}\lambda^{1/4}},\label{eq:nlo-norm-A}
\end{align}
then \refeq{eq:nlo-norm-comrels03} becomes  
\begin{align}
\nonumber [\hat{X}(t,\sigma),n_t\hat{\mathcal{P}}^t(t,\sigma^\prime)] &=\frac{i}{\sqrt{g|_\Sigma}}\int\limits_{-\infty}^\infty\frac{d\omega}{2\pi}\left[e^{i\omega(\sigma-\sigma^\prime)}+e^{i\omega(\sigma+\sigma^\prime)}\right] \\
&= \frac{i}{\sqrt{g|_\Sigma}}\left[\delta(\sigma-\sigma^\prime)+\delta(\sigma+\sigma^\prime)\right] = i\frac{\delta(\sigma-\sigma^\prime)}{\sqrt{g|_\Sigma}}, \label{eq:nlo-norm-comrels04}
\end{align}
where in the last line we dropped the second Dirac delta function since $\sigma,\sigma^\prime\in[0,\sigma_f]$. What we have shown is that given the normalisation ansatz \refeq{eq:nlo-norm-A}, \refeq{eq:nlo-norm-comrels04} reproduces \refeq{eq:nlo-norm-comrels01}.
\par 
We denote the operator corresponding to the position of the free endpoint by
\begin{align}
\hat{X}_\text{End}(t;a):=\hat{X}(t,\sigma_f-at).
\end{align} 
According to \cite{Birrell:1982ix}, the transverse fluctuations are excited at the Hawking temperature by the Hawking radiation emitted by the black hole horizon, given by \refeq{eq:gen-hawkingtemp}, where, for our semi-classical treatment, the excitations are purely thermal \cite{Atmaja:2010uu,deBoer:2008gu}. These excitations are described by the Bose-Einstein distribution \cite{Atmaja:2010uu,deBoer:2008gu} 
\begin{align}
\avg{\hat{a}_\omega^\dagger \hat{a}_{\omega^\prime}}=\frac{2\pi\delta(\omega-\omega^\prime)}{e^{\beta\omega}-1}. \label{eq:nlo-thermal}
\end{align} 
Using \refeq{eq:nlo-thermal} we can compute the following normal ordered correlator
\begin{align}
\nonumber &\avg{\normalorder{\hat{X}_\text{End}(t_1;a)\hat{X}_\text{End}(t_2;a)}} \\
\nonumber &= \frac{\beta^2}{8\pi^2\sqrt{\lambda}}\int\limits_{0}^{\infty}\frac{d\omega_1d\omega_2}{\sqrt{\omega_1\omega_2}}\left(f_{\omega_1}(\sigma_f-at_1)f_{\omega_2}^\ast(\sigma_f-at_2)e^{-i\omega_1t_1+i\omega_2t_2}\frac{\avg{\hat{a}_{\omega_1}^\dagger \hat{a}_{\omega_2}}}{2\pi}\right. \\
\nonumber &\hspace{3.8cm}\left.+f_{\omega_1}^\ast(\sigma_f-at_1)f_{\omega_2}(\sigma_f-at_2)e^{-i\omega_2t_2+i\omega_1t_1}\frac{\avg{\hat{a}_{\omega_2}^\dagger \hat{a}_{\omega_1}}}{2\pi}\right)
\\ &= \frac{\beta^2}{4\pi^2\sqrt{\lambda}}\int\limits_{0}^{\infty}\frac{d\omega}{\omega}\frac{1}{e^{\beta\omega}-1}\text{Re}\left(f_{\omega}(\sigma_f-t_1)f_{\omega}^\ast(\sigma_f-t_2)e^{-i\omega(t_1-t_2)}\right) \label{eq:nlo-correlator}
\end{align}
where we have introduced the normal ordering operator $\normalorder{\hat{a}_\omega^\dagger \hat{a}_{\omega^\prime}}=\normalorder{\hat{a}_{\omega^\prime}\hat{a}_\omega^\dagger}=\hat{a}_\omega^\dagger \hat{a}_{\omega^\prime}$ as a regulator to avoid otherwise present logarithmic UV divergences \cite{deBoer:2008gu}. 

Following \cite{deBoer:2008gu}, we may compute the average transverse distance squared travelled by the endpoint of our falling string, 
\begin{align}
\nonumber s^2(t;a):&=\avg{\normalorder{(\hat{X}_\text{End}(t;a)-\hat{X}_\text{End}(0;a))^2}} \\
\nonumber &= \avg{\normalorder{\hat{X}^2_\text{End}(t;a)}}+\avg{\normalorder{\hat{X}^2_\text{End}(0;a)}}-2\avg{\normalorder{\hat{X}_\text{End}(t;a)\hat{X}_\text{End}(0;a)}} \\
&= \frac{\beta^2}{4\pi^2\sqrt{\lambda}}\int\limits_{0}^{\infty}\frac{d\omega}{\omega}\frac{1}{e^{\beta\omega}-1}\left|f_{\omega}(\sigma_f-at)-f_{\omega}(\sigma_f)e^{i\omega t}\right|^2.
\label{eq:nlo-s}
\end{align}

For $a=0$, one may show that \refeq{eq:nlo-s} reduces to Eq.~3.5 of \cite{deBoer:2008gu}, 
\begin{align}
\nonumber s^2(t,0) &= \frac{\beta^2}{\pi^2\sqrt{\lambda}}\int\limits_{0}^{\infty}\frac{d\omega}{\omega}\frac{\sin^2\left(\frac{\omega t}{2}\right)}{e^{\beta\omega}-1}\left|f_{\omega}(\sigma_f)\right|^2\\
&= \frac{4 \beta^2}{\pi^2\sqrt{\lambda}}\int\limits_{0}^{\infty}\frac{d\omega}{\omega}\frac{\sin^2\left(\frac{\omega t}{2}\right)}{e^{\beta\omega}-1}\frac{1+\nu^2}{1+\nu^2\tilde{r}_0^2}, \label{eq:nlo-s0}
\end{align}
a necessary and non-trivial consistency check on our work.

Recall that we previously made the observation that the leading order solution for a string whose endpoint falls at a fraction $a$ of the local speed of light may be thought of as an observer travelling down the leading order static stretched string solution at that same fraction $a$ of the local speed of light.  We find here that, similarly, the semi-classical next-to-leading order transverse motion experienced by the endpoint of a string that falls at a fraction $a$ of the local speed of light is \emph{exactly the same} as the transverse motion experienced by an observer travelling down the semi-classical next-to-leading order static stretched string solution at that same fraction $a$ of the local speed of light.  

\subsection{Small Virtuality\label{sec:small}}
The small virtuality limit consists of $\ell_0\ll r_H$\footnote{For the phenomenologically relevant case of $\lambda \sim \mathcal{O} (10)$ and $T\sim0.5$ GeV, there is a large region of $\ell_0$ parameter space for which $L/\lambda^{1/4} \sim 0.5 \ll \ell_0 \ll r_H=4\pi L^2T/(d-1)\sim7.5$.}. In this case,
\begin{align*}
\coth^2\left(\frac{r_H}{L^2}({r_s}_\ast+\sigma)\right) \approx 1, 
\end{align*} 
and \refeq{eq:nlo-eomsig} reduces to the wave equation
\begin{align}
(\partial_t^2-\partial_\sigma^2)X=0. \label{eq:small-eom}
\end{align}
\refeq{eq:small-eom} is solved by the linearly independent out-going (progressive) and infalling (regressive) modes calculated in \refeq{eq:nlo-pm} which reduce to the near-horizon boundary conditions given by \refeq{eq:nlo-bcs}; namely
\begin{align}
f_\omega^{(\pm)}(\sigma)= e^{\pm i\omega({r_s}_\ast+\sigma))}. \label{eq:small-bcs}
\end{align}
We can calculate the coefficient $B_\omega$ according to \refeq{eq:nlo-B} as
\begin{align}
B_\omega = e^{i2\omega({r_s}_\ast+\sigma_f)}. \label{eq:small-B}
\end{align}
Putting all the pieces together from \refeq{eq:small-bcs} and \refeq{eq:small-B}, $f_\omega$ can be written as
\begin{align}
f_\omega(\sigma) = 2e^{i\omega({r_s}_\ast+\sigma_f)}\cos(\omega(\sigma-\sigma_f)). \label{eq:small-f}
\end{align}
Substituting \refeq{eq:small-f} into \refeq{eq:nlo-s} yields
\begin{align}
\nonumber s^2_\text{small}(t;a)&=\frac{\beta^2}{\pi^2\sqrt{\lambda}}\int\limits_{0}^{\infty}\frac{d\omega}{\omega}\frac{1}{e^{\beta\omega}-1}\left(1+\cos^2(a \omega t)-2\cos(a\omega t)\cos(\omega t)\right)\\
&=\frac{\beta^2}{4\pi^2\sqrt{\lambda}}\ln\left(\frac{2 a \beta ^3 \sinh ^2\left(\frac{\pi  (a+1) t}{\beta }\right) \sinh ^2\left(\frac{\pi(a-1)t}{\beta }\right)  \text{csch}\left(\frac{2 \pi  a t}{\beta }\right)}{\pi ^3 \left(a^2-1\right)^2 t^3}\right),\label{eq:small-s}
\end{align}
where in the last line we used \refeq{eq:gen-hawkingtemp}. Notice that \refeq{eq:small-s} is independent of the IR regulator, as it must be. For later convenience we shall refer to \refeq{eq:small-s} as the small virtuality result, hence the subscript. 

One may readily compute the two limiting cases of interest for $s^2_\text{small}(t;a)$ for the speed at which the endpoint falls, $a=0$, which corresponds to the static string (on-shell heavy quark), and $a=1$, which corresponds to the freely falling limp noodle (off-shell light quark).  For $a=0$
\begin{align}
s^2_\text{small}(t;0)=\frac{\beta ^2}{\pi ^2 \sqrt{\lambda}}\ln \left(\frac{\beta  \sinh \left(\frac{\pi  t}{\beta }\right)}{\pi  t}\right), \label{eq:small-s0}
\end{align}
and for $a=1$
\begin{align}
s^2_\text{small}(t;1)=\frac{ \beta ^2}{4 \pi ^2 \sqrt{\lambda}}\ln \left(\frac{\beta  \sinh \left(\frac{2 \pi  t}{\beta }\right)}{2 \pi  t}\right).
\end{align}

From the full solution $s^2_\text{small}(t;a)$, \refeq{eq:small-s}, we see that $\beta$ sets a natural cross-over time scale between early and late time dynamics. Expanding \refeq{eq:small-s} in powers of $t/\beta$ we obtain the early time behaviour 
\begin{align}
s_\text{small}^2(t;a)&\xrightarrow{t\ll\beta} \frac{t^2}{6 \sqrt{\lambda}} + \mathcal{O}\big((t/\beta)^4\big) \label{eq:small-s-early}.
\end{align}
One sees that the early time dynamics is ballistic, $s^2(t) \sim t^2$, and that the coefficient of the $t^2$ term is independent of $a$, the speed at which the free endpoint is allowed to fall.  The latter is not a surprise: we expect that at early enough times, $t\ll\beta$, the string endpoint ``doesn't know'' whether or not it is free to fall.  It is interesting that the temperature alone sets the timescale for this information propagation to the string endpoint; there is no dependence on the initial length of the string, $\ell_0$ (i.e.\ virtuality of the light quark). 

Although $t\in[0,\sigma_f]$ is bounded, mathematically the late time behaviour can be obtained by expanding \refeq{eq:small-s} in powers of $\beta/t$
\begin{align}
s_\text{small}^2(t;a)&\xrightarrow{\beta\ll t} \frac{\beta t}{\pi  \sqrt{\lambda}}\left(1-\frac{a}{2}\right) + \frac{\beta^2}{4 \pi^2 \sqrt{\lambda}}\left\{
\begin{array}{ll}
4\ln\left(\frac{\beta}{2\pi t}\right)&, \text{if}~a=0\\
\ln\left(\frac{a \beta ^3}{4 \pi ^3 \left(a^2-1\right)^2 t^3}\right)&,\text{if}~0<a<1\\
\ln\left(\frac{\beta}{4\pi t}\right)&,\text{if}~a=1
\end{array}
\right\} + \mathcal{O}(1). \label{eq:small-s-late}
\end{align}
The late time behaviour is therefore diffusive at leading order, $s^2(t)\sim 2D(a)t$, with diffusion coefficient
\begin{align}
D(a):= \frac{\beta}{2\pi  \sqrt{\lambda}}\left(1-\frac{a}{2}\right), \label{eq:small-D}
\end{align}
but with sub-diffusive log corrections.

The diffusion coefficient, \refeq{eq:small-D}, consists of two factors: The first factor $\frac{\beta}{2\pi\sqrt{\lambda}}$ is precisely the diffusion coefficient for a stretched string (on-shell heavy quark) given in Eq.~3.9 of \cite{deBoer:2008gu}. The second factor $1-\frac{a}{2}$ is the previously encountered fraction of the area of the parameter-space whose image is a one-to-one map onto the entire world-sheet in the target space. 

We therefore find that a small virtuality light quark initially at rest in a strongly-coupled plasma has exactly one half the diffusion coefficient of a massive heavy quark.

\subsection{Arbitrary Virtuality\label{sec:arb}}

\subsubsection{Asymptotic Early Time Dynamics\label{sec:arb-early}}
For asymptotically early times, we expand the time-dependent component of the integrand in \refeq{eq:nlo-s} in powers of $t/\beta$
\begin{align}
\nonumber |f_\omega(\sigma_f-at)-f_\omega(\sigma_f)e^{i\omega t}|^2\xrightarrow{t\ll\beta}&|f_\omega(\sigma_f)|^2\left(\omega^2t^2+\mathcal{O}(t/\beta)^4\right)\\
=&\frac{16\pi^2\nu^2t^2}{\beta^2}\frac{1+\nu^2}{1+\nu^2 \tilde{r}_0^2}+\mathcal{O}(t/\beta)^4, \label{eq:arb-early-exp}
\end{align}
where we remember that 
\begin{align*}
 \tilde{r}_0=\frac{r_s+\ell_0}{r_H}.
\end{align*}
Substituting the LO contribution from \refeq{eq:arb-early-exp} into \refeq{eq:nlo-s}, we find
\begin{align}
\nonumber \left.s^2(t;a)\right|_{t\ll\beta}=&\frac{4t^2}{\sqrt{\lambda}}\int\limits_{0}^{\infty}d\nu\frac{\nu}{e^{2\pi\nu}-1}\frac{1+\nu^2}{1+\nu^2 \tilde{r}_0^2}\\
=&\frac{t^2}{6 \tilde{r}_0^4\sqrt{\lambda}}\left( \tilde{r}_0^2-6( \tilde{r}_0^2-1)\left(-2\gamma_E-\pi\cot\left(\frac{\pi }{ \tilde{r}_0}\right)+H_{-1/ \tilde{r}_0}+H_{1/ \tilde{r}_0}+2\ln( \tilde{r}_0)\right)\right), \label{eq:arb-early-s}
\end{align}
where $\gamma_{E}$ is the Euler-Mascheroni constant and $H_{\mp1/\tilde{r}_0}$ are harmonic numbers. The integral leading to \refeq{eq:arb-early-s} was evaluated by recognising that the initial expression is decomposable into the sum of two terms that are the second derivatives with respect to $k$ (evaluated at $k=0$) of the integrals in Eq.~B.2 of \cite{deBoer:2008gu}. 

Just as in the small initial length (virtuality) case described in \refsec{sec:small}, it is the temperature alone that sets the scale for early vs.\ late time behaviour, independent of the initial string length $\ell_0$.  Similarly, the early time dynamics is ballistic, $s^2(t)\sim t^2$, with a coefficient that is independent of $a$, the speed that the free string endpoint falls.

As two consistency checks of our early time, arbitrary initial length (virtuality) result, \refeq{eq:arb-early-s}, we can consider the small and large virtuality limits; i.e.~$\tilde{r}_0\to 1$ and $\tilde{r}_0\to\infty$, respectively. In the small virtuality limit, \refeq{eq:arb-early-s} reduces to
\begin{align}
\left.s^2(t;a)\right|_{t\ll\beta}\xrightarrow{ \tilde{r}_0\to1+\epsilon}\frac{ t^2}{6 \sqrt{\lambda}}+\mathcal{O}(\epsilon),\label{eq:arb-early-s-small}
\end{align}
which is precisely the early time behaviour of the small virtuality result we already found, \refeq{eq:small-s-early}. For large virtualities, \refeq{eq:arb-early-s} becomes
\begin{align}
\left.s^2(t;a)\right|_{t\ll\beta}\xrightarrow{ \tilde{r}_0\to\infty}\frac{t^2}{ \tilde{r}_0 \sqrt{\lambda}}+\mathcal{O}\big((1/ \tilde{r}_0)^2\big), \label{eq:arb-early-s-large}
\end{align}
which we recognise as precisely the early time behaviour observed in Eq.~3.6 of \cite{deBoer:2008gu}.

\subsubsection{Asymptotic Late Time Dynamics\label{sec:arb-late}}
For asymptotically late times, it is instructive to define the following dimensionless quantities: $z:=\omega t$ and $x:=\frac{\beta}{t}$, where we shall take $x\ll1$. Rewriting \refeq{eq:nlo-s} in terms of $z$ and $x$ we obtain
\begin{align}
s^2(t;a) = \frac{\beta^2}{4\pi^2\sqrt{\lambda}}\int\limits_{0}^{\infty}\frac{dz}{z}\frac{1}{e^{z x}-1}\left|f_{z/t}(\sigma_f-a\beta/x)-f_{z/t}(\sigma_f)e^{iz}\right|^2. \label{eq:arb-late-sfull}
\end{align}
Expanding the same component of the integrand as was performed previously, but in powers of $\beta/t=:x$ yields
\begin{align}
\left|f_{z/t}(\sigma_f-a\beta/x)-f_{z/t}(\sigma_f)e^{iz}\right|^2\xrightarrow{x\ll 1}4\left(1+\cos^2(a \omega t)-2\cos(a\omega t)\cos(\omega t)\right)+\mathcal{O}(x).\label{eq:arb-late-exp}
\end{align}
Substituting the LO contribution from \refeq{eq:arb-late-exp} into \refeq{eq:arb-late-sfull}, we obtain
\begin{align}
\nonumber \left.s^2(t;a)\right|_{t\gg\beta}=&\frac{\beta^2}{\pi^2\sqrt{\lambda}}\int\limits_{0}^{\infty}\frac{dz}{z}\frac{1}{e^{z x}-1}\left(1+\cos^2(a z)-2\cos(az)\cos(z)\right)\\
\nonumber =&\frac{\beta^2}{\pi^2\sqrt{\lambda}}\int\limits_{0}^{\infty}\frac{d\omega}{\omega}\frac{1}{e^{\omega\beta}-1}\left(1+\cos^2(a \omega t)-2\cos(a\omega t)\cos(\omega t)\right)\\
=& s^2_\text{small}(t;a), \label{eq:arb-late-s}
\end{align}
where we recognise the second line \refeq{eq:arb-late-s} as being the first line of \refeq{eq:small-s}. 

Quite remarkably, we have uncovered universal behaviour: the diffusive asymptotic late time behaviour of $s^2(t;a)$ is fully encoded in the information-dense $s^2_\text{small}(t;a)$ and is, in particular, independent of the initial length of the string.  It turns out that we can solve for $s^2_\text{small}(t;a)$ in arbitrary dimension $d$, and, thus, we are able to extract the late time behaviour (i.e.\ diffusion coefficient) for strings of arbitrary length (quarks of arbitrary mass/virtuality) in arbitrary spatial dimensions, including the phenomenologically relevant case of $AdS_5$.

\subsection{\texorpdfstring{$AdS_d$}{AdS\_d}-Schwarzschild for \texorpdfstring{$d\ge3$}{d>=3}\label{sec:gen}}
Suppose we want to determine $s^2_\text{small}(t;a,d)$ - the small virtuality result in $AdS_d$-Schwarz\-schild for $d\ge3$. For sufficiently small virtualities, we are interested in the near-horizon geometry of the black-brane. To this end, let $r=(1+\varepsilon)r_H$. Expanding (and truncating) each term in the $AdS_d$ metric, \refeq{eq:gen-metric}, to lowest non-vanishing order in $\varepsilon$, we obtain
\begin{align}
\nonumber ds_d &= \frac{r_H^2}{L^2}\left(-(d-1)\varepsilon dt^2 + d\bm{x}_{d-2}^2\right) + \frac{L^2}{r_H^2}\frac{d\varepsilon^2}{(d-1)\varepsilon}\\
&= \frac{r_H^2}{L^2}(d-1)\exp\left(\frac{r_H^2}{L^2}(d-1)\varepsilon_\ast\right)(-dt^2+d\varepsilon_\ast^2)+ \frac{r_H^2}{L^2}d\bm{x}_{d-2}^2, \label{eq:gen-metric-iso}
\end{align}
where 
\begin{align}
\begin{split}
\varepsilon_\ast &:=\frac{L^2}{r_H^2}\frac{\ln(\varepsilon)}{(d-1)}\\
\varepsilon &:= \exp\left(\frac{r_H^2}{L^2}(d-1)\varepsilon_\ast\right),
\end{split} \label{eq:gen-tortoisecoord-iso}
\end{align}
are the so-called tortoise and inverse-tortoise coordinates. \refeq{eq:gen-tortoisecoord-iso} is consistent with expanding \refeq{eq:gen-tortoisecoord} for $r=(1+\varepsilon)r_H$ about $\varepsilon=0$ up to the addition of a physically meaningless constant. Therefore we are able to invert the tortoise coordinate in the near horizon limit for arbitrary dimensions.
\par
Without knowledge of the LO limp noodle dynamics in $AdS_d$-Schwarzschild, we can consider small virtuality fluctuations in some transverse direction of $AdS_d$-Schwarzschild, which we shall denote by $X:[0,\sigma_f]^2\to\reals$. Since the metric in \refeq{eq:gen-metric-iso} is independent of $\epsilon$ in the transverse directions, the equation of motion for $X$ (according to \refeq{eq:ng-eom}) is the familiar wave equation given by \refeq{eq:small-eom}.
Repeating the computation in \refsec{sec:small}, we obtain
\begin{align}
s^2_\text{small}(t;a,d)
&=\frac{1}{\sqrt{\lambda}}\left(\frac{(d-1)\beta}{4\pi}\right)^2\ln\left(\frac{2 a \beta ^3 \sinh ^2\left(\frac{\pi  (a+1) t}{\beta }\right) \sinh ^2\left(\frac{\pi(a-1)t}{\beta }\right)  \text{csch}\left(\frac{2 \pi  a t}{\beta }\right)}{\pi ^3 \left(a^2-1\right)^2 t^3}\right),\label{eq:gen-small-s}
\end{align}
which reduces to our previous result for $d=3$, \refeq{eq:small-s}. Expanding \refeq{eq:gen-small-s} in powers of $\beta/t$, we can again extract the diffusion coefficient
\begin{align}
D(a,d):= \frac{(d-1)^2\beta}{8\pi \sqrt{\lambda}}\left(1-\frac{a}{2}\right).
\end{align}
For $d=5$ we reproduce the known low velocity heavy quark ($a=0$) diffusion coefficient \cite{Gubser:2006nz,CasalderreySolana:2006rq,Herzog:2006gh}.

\section{Discussion and Outlook\label{sec:concl}}
In this paper we computed the average transverse distance squared travelled by the free endpoint of a string initially at rest whose other endpoint is fixed to a black hole horizon in $AdS_3$-Schwarzschild, $s^2(t,a)$, as a function of time $t$ and as a function of the fraction $a$ of the local speed of light that the free endpoint of the string is allowed to fall in the third dimension.

We analytically evaluated in closed form $s_\text{small}^2(t,a)$ for small initial string lengths as compared to the Hawking radius of the black hole, $\ell_0\ll r_H$ .  We found that the temperature alone sets the scale for early and late time behaviour, independent of the initial length of the string $\ell_0$.  At early times, $t\ll\beta$, the motion is ballistic, with $s^2(t,a)\sim t^2$, independent of the string endpoint falling speed $a$.  At late times, $t\gg\beta$, the motion is to leading order in time diffusive, with $s^2(t,a)\sim t$, with sub-diffusive log corrections.  In contradistinction to Fick's Law, in which $s^2(t)\sim t$ at late times in any number of Euclidean dimensions, the presence of the anomalous diffusion may indicate a fractal nature to the strongly-coupled plasma or that the plasma should be considered a disordered medium \cite{Havlin}.  Similar anomalous diffusion has been observed in ultra-cold atoms and is characteristic of physical situations in which there is no length scale \cite{Sagi}.  

Next, we derived an integral expression for $s^2(t,a)$ for arbitrary initial string lengths $\ell_0$.  Again, the temperature alone sets the scale that separates early from late time dynamics, independent of the initial string length $\ell_0$.  As in the small initial string length case, $\ell_0\ll r_H$, the early time behaviour is ballistic, $s^2(t,a)\sim t^2$, with a proportionality constant that is now a function of $\ell_0$ (which becomes independent of $\ell_0$ again as $\ell_0$ becomes small compared to $r_H$).  Interestingly, the late time behaviour for a string of arbitrary initial length is identical to that of a string of small initial length, $s^2(t\gg\beta,a) = s_\text{small}^2(t,a)$, where late time is still determined solely by the temperature of the plasma (i.e.\ is independent of initial string length).

Motivated by the latter late time universal behaviour, we generalized our derivation of $s_\text{small}^2(t,a)$ to $AdS_d$-Schwarzschild metrics in $d$ dimensions, yielding $s^2_\text{small}(t,a,d) = \frac{(d-1)^2}4s^2_\text{small}(t,a)$.  Thus the qualitative early and late time behaviour, ballistic and diffusive, respectively, is the same for all initially small length strings in any number of dimensions.

Since the late time behaviour is universal, we extracted the diffusion coefficient $D(a,d)$ from $s^2(t\gg\beta,a,d)=2D(a,d)t$,

\begin{align}
D(a,d):= \frac{(d-1)^2\beta}{8\pi \sqrt{\lambda}}\left(1-\frac{a}{2}\right) \xrightarrow{d\rightarrow5} \left\{
\begin{array}{ll}
	2\beta/\pi\sqrt{\lambda}, & \text{ heavy quark } (a=0) \\[5pt]
	\beta/\pi\sqrt{\lambda}, & \text{ light quark } (a=1),
\end{array}
\right.
\end{align}
where we explicitly show the result for the phenomenologically relevant case of $d=5$.  Surprisingly, in any number of dimensions, the diffusion coefficient for a light quark, $a=1$, is precisely one half that of a heavy quark, $a=0$.  

Our investigation of the limp noodle was, however, originally birthed out of a desire to understand the fluctuations in the motion of a fast moving light or heavy quark.  We now argue that our above results actually give us access to precise information on the transverse fluctuations in the motion of a fast moving string in arbitrary numbers of dimensions; in particular, our work gives the transverse fluctuations as a function of time for a light or heavy quark propagating in a strongly-coupled plasma in four spacetime dimensions.

In their paper, Chesler et al.\ \cite{Chesler:2008uy} showed that falling string endpoints follow trajectories that can be approximated by null geodesics. Suppose that such a null geodesic lies in the $r,x^1$-plane. Focusing on a single string endpoint, if one were to project such a null geodesic onto the transverse $r,\bm{x}_\perp$-plane (where $\bm{x}_\perp = (x^2,\cdots,x^{d-2})$), then one could use the tortoise coordinate defined in \refeq{eq:lo-ads-tortoisecoord} to extract $a(r)$---the fraction of the speed of light at which the free endpoint falls in $\reals^{1,1}$. Once we know $a(r)$, we know the diffusion coefficient $D(a(r),d)$ for the transverse fluctuations of the falling string solution. 

\begin{figure}[!tbp]
\centering
\includegraphics[width=\linewidth]{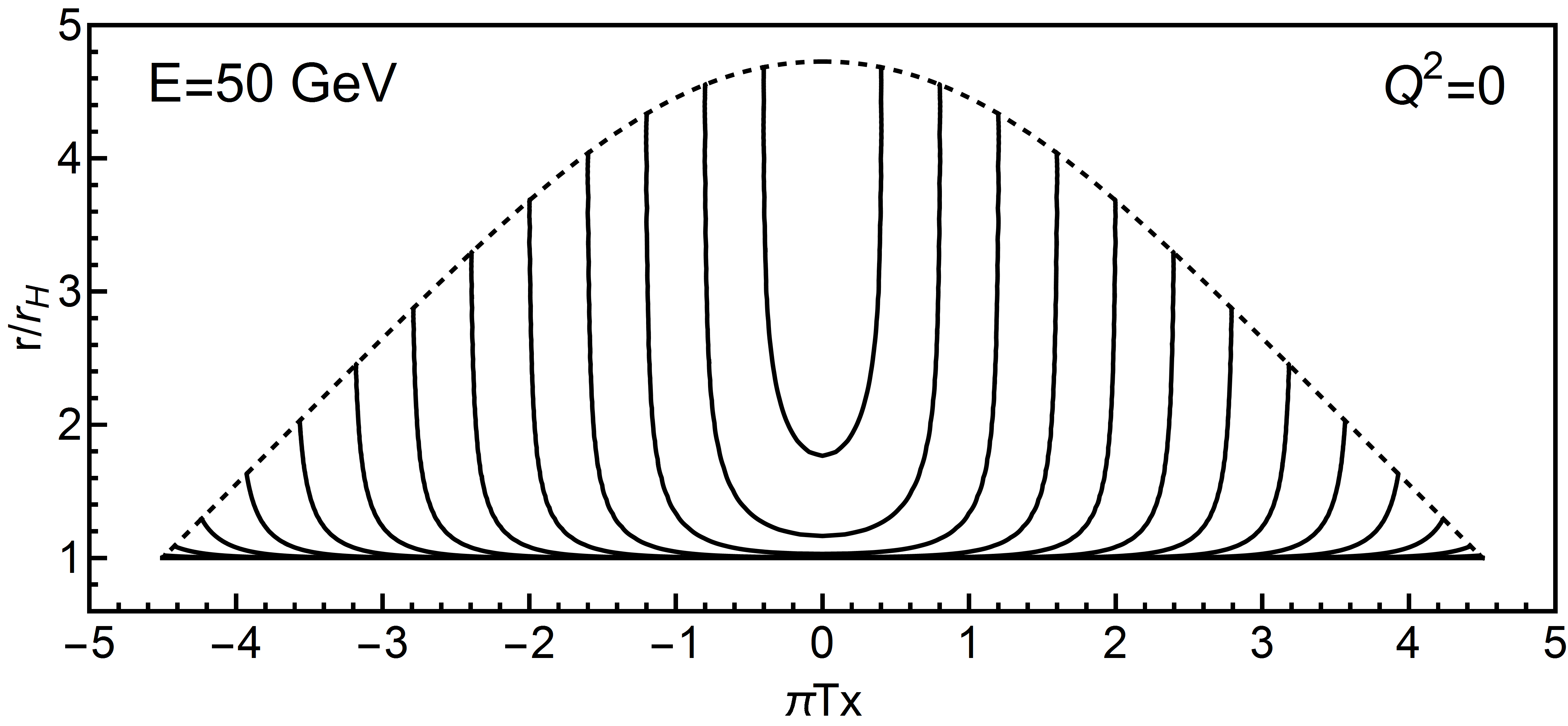}
\caption{(Colour online) Snapshots in time of a typical falling string profile (black curve) of Chesler et al.\ \protect\cite{Chesler:2008uy}. The falling string nucleates at $x=0$ as a point and evolves to an extended object with the endpoints falling towards the stretched horizon at $r\sim r_H$ along trajectories (dashed black curve) approximated by null geodesics. Initially the transverse fluctuations are characterised by $a=0$ while close to the stretched horizon $a=1$.}
\label{fig:fall-profile}
\end{figure}

\reffig{fig:fall-profile} displays a typical Chesler et al.\ falling string solution \cite{Chesler:2008uy} for a 50 GeV quark with $Q^2 = 0$ in a $T = 350$ MeV plasma. One can see from the figure that $a$ is a monotonically increasing function of time (or, equivalently, distance travelled by the string endpoint). Thus the quark experiences its greatest transverse fluctuations at the initial time, and the fluctuations decrease monotonically with time.  This decrease in transverse fluctuations with time/distance is consistent with the pQCD result of angularly ordered gluon emissions from an off-shell parton in vacuum \cite{Mueller:1981ex,Ermolaev:1981cm} and in contradistinction to the anti-angular ordering observed in medium \cite{MehtarTani:2010ma}. More quantitatively, the transverse fluctuations of the falling string solution cause the string to diffuse half as much at late times as at early times. The precise interpretation of this factor of a half is unclear and is the topic of future work.

The transverse momentum squared per unit length picked up by a fast moving parton, the transport coefficient $\hat{q}$, is an important property of a plasma. $\hat{q}$ can be related to a number of interesting observables---for example the angular decorrelation of jets in heavy ion collisions \cite{Renk:2006pk}---and can be connected, in the weak coupling limit, to the radiative energy loss of the parton propagating through the plasma \cite{Majumder:2010qh,Burke:2013yra}.  Two general approaches have been used to compute $\hat{q}$ in the strong-coupling limit, yielding two different results.  

In \cite{Gubser:2006nz}, Gubser found $\hat{q}_{\text{Gubser}}=2\pi\sqrt{\lambda}\gamma^{1/2}T^3/v$ by considering the motion of a constant velocity infinitely massive heavy quark.  As pointed out in the original work \cite{Gubser:2006nz} as well as in \cite{CasalderreySolana:2007qw}, the heavy flavor setup used in \cite{Gubser:2006nz} to compute $\qhat$ breaks down for heavy quark velocities larger than a speed limit, $\gamma_\text{crit}=(1+2M_Q/T\surd\lambda)^2$, above which the heavy quark should be treated as an off-shell effectively massless quark.  Additionally, one might object to the infinite time setup used, which leads to an infinite energy in the string; it's possible that this infinite energy stored in the string then ``wags the dog,'' leading to the rapid increase in the size of fluctuations with $\gamma$.  This interpretation is supported by \cite{Giecold:2009cg}, in which it was emphasized that these fluctuations that do not obey the usual fluctuation-dissipation relations are due to the presence of the \emph{induced} worldsheet horizon resulting from the enforced constant velocity of the heavy quark and not the horizon in the spacetime that represents the thermal medium. 

\begin{figure*}[!tbp]
\hspace*{\fill}%
\begin{subfigure}[t]{0.48\textwidth}
\centering
\includegraphics[width=\columnwidth]{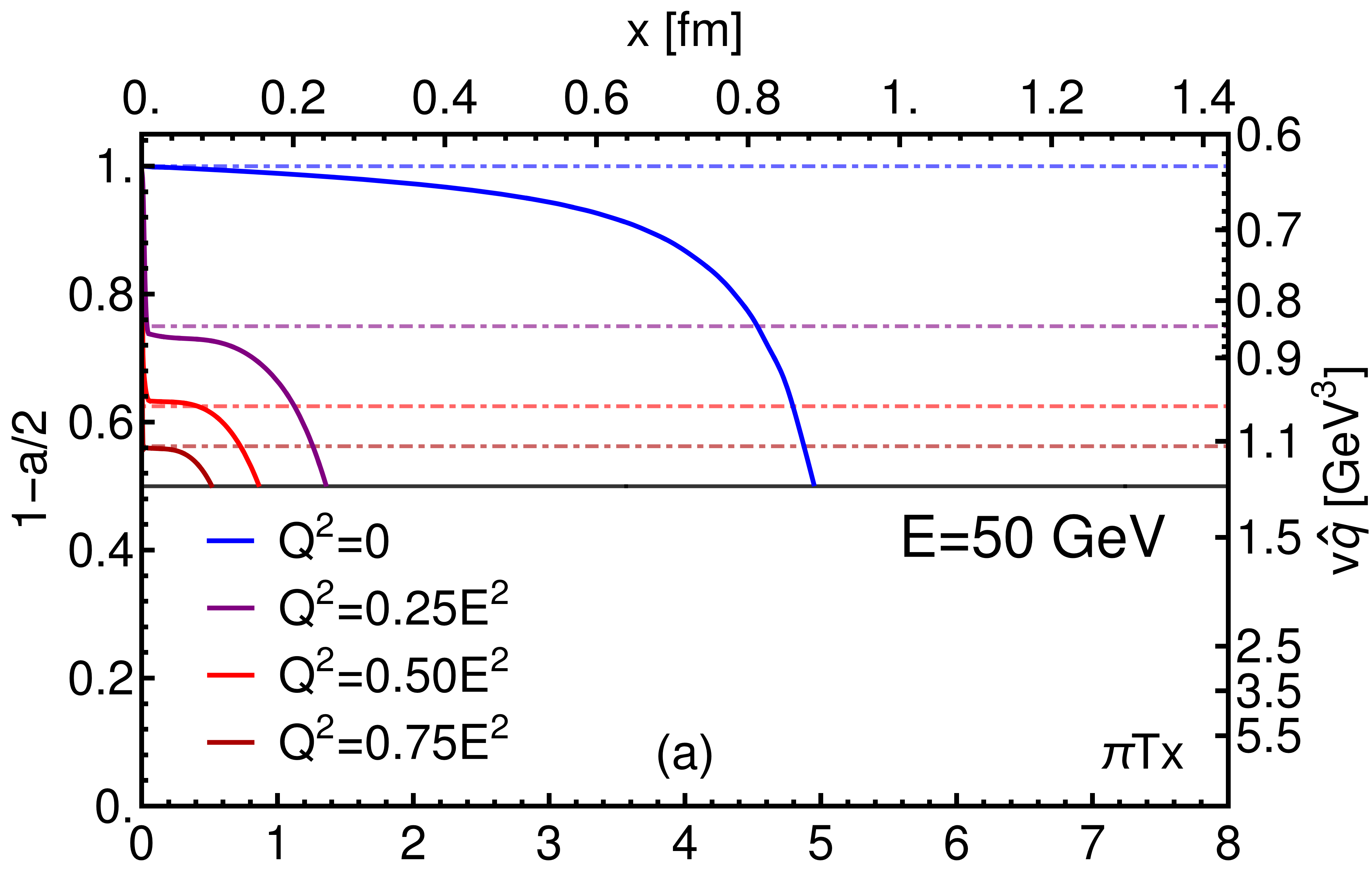}
\end{subfigure}\hspace{.15in}
\begin{subfigure}[t]{0.48\textwidth}
\centering
\includegraphics[width=\columnwidth]{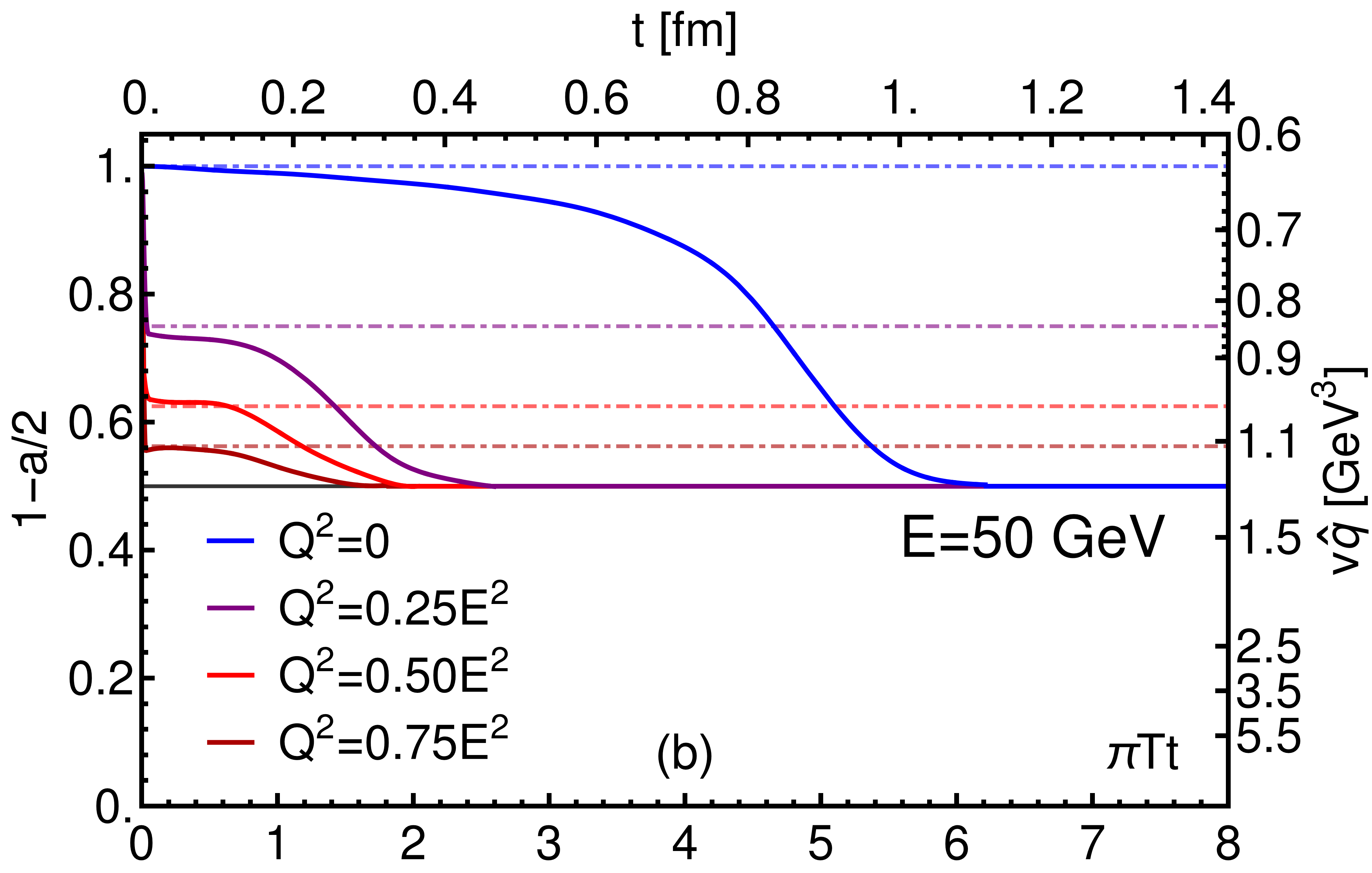}
\end{subfigure}
\begin{subfigure}[t]{0.48\textwidth}
\centering
\includegraphics[width=\columnwidth]{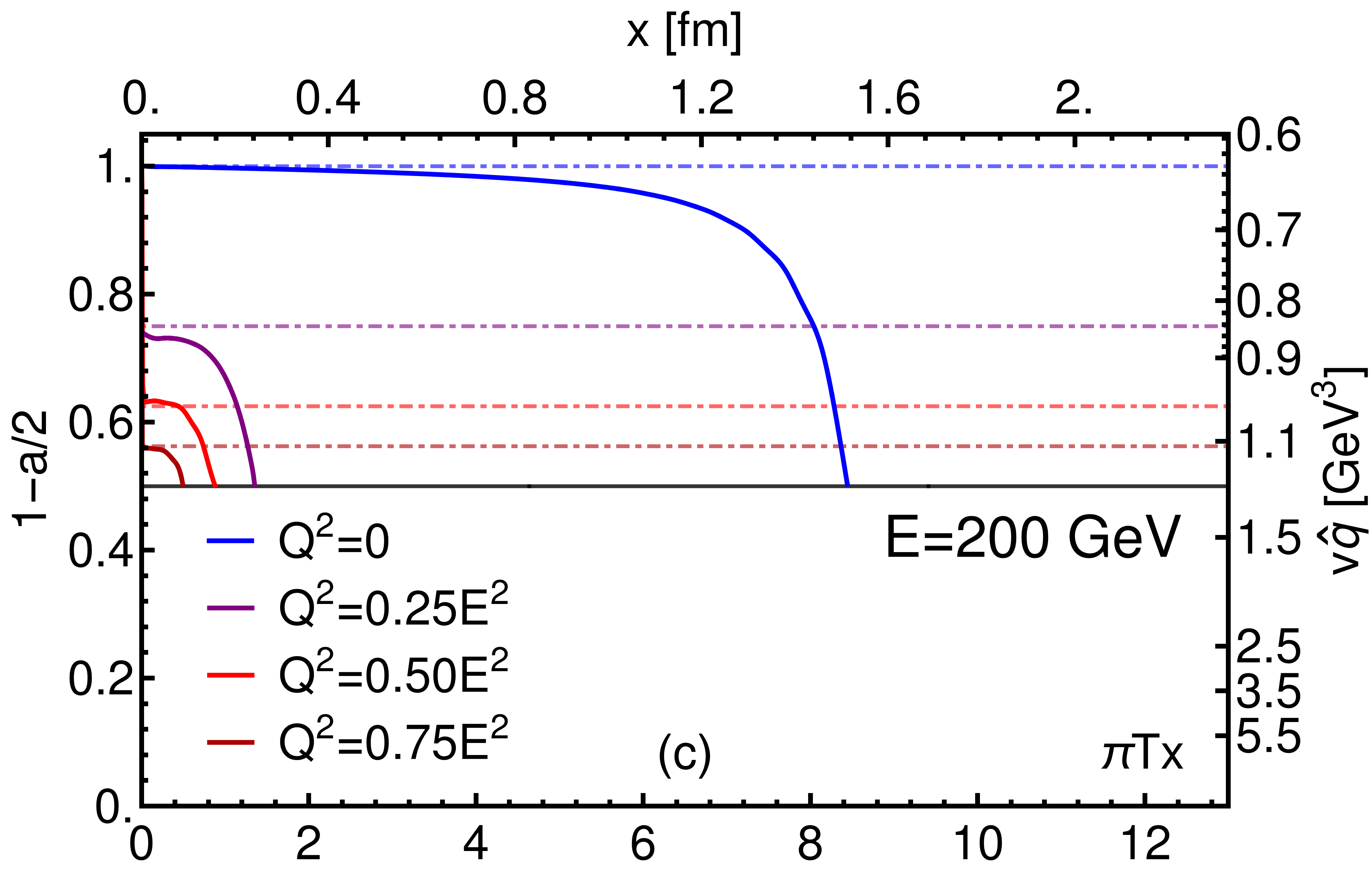}
\end{subfigure}\hspace{.15in}
\begin{subfigure}[t]{0.48\textwidth}
\centering
\includegraphics[width=\columnwidth]{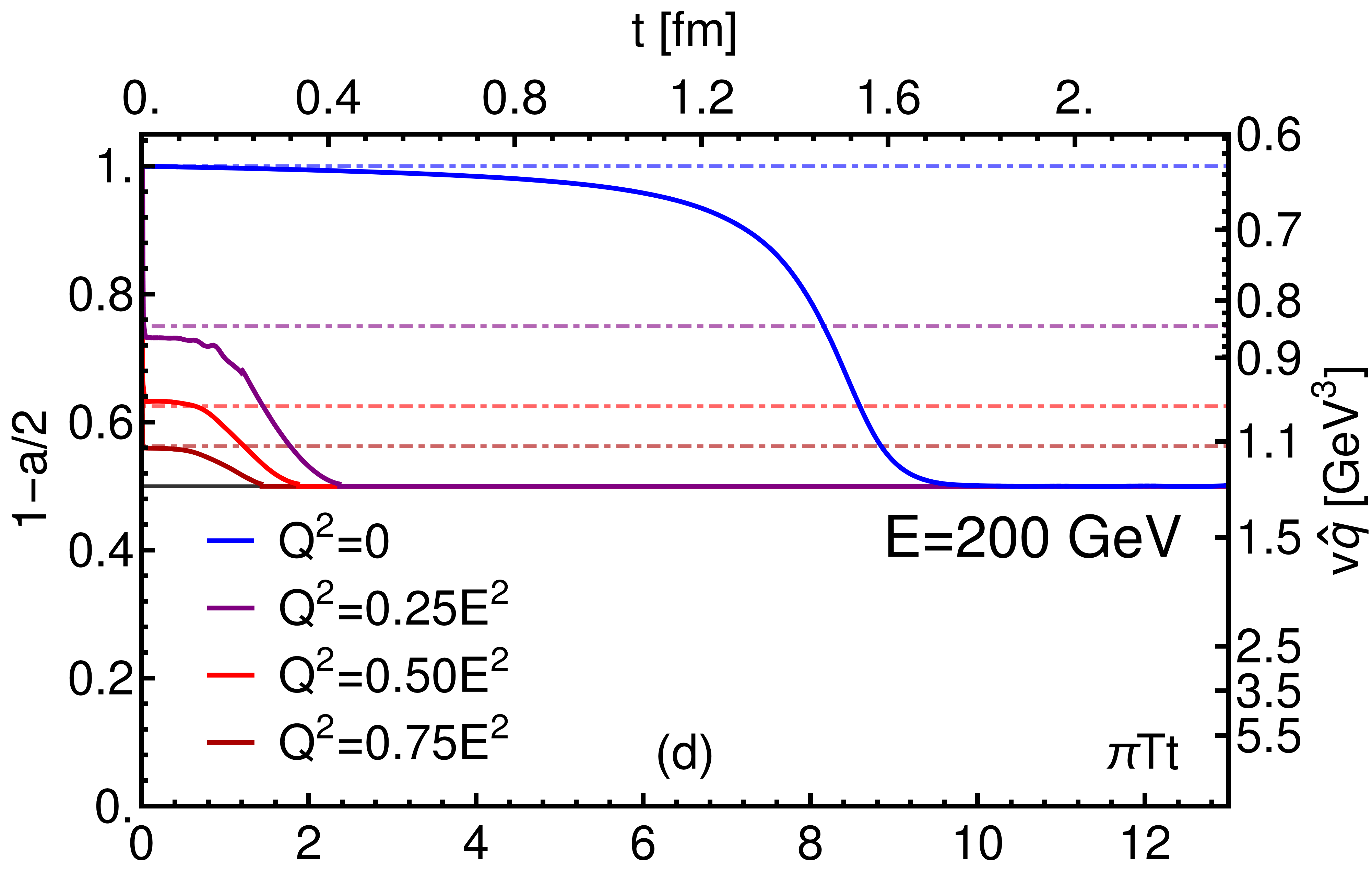}
\end{subfigure}
\hspace*{\fill}
\caption{(Colour online) $1-a/2\propto1/(v\times\hat{q})$ plotted as a function of $x$ (left column) and $t$ (right column) for an $E =  50$ GeV quark (top row) and $E = 200$ GeV quark (bottom row) for various virtualities in a $T = 350$ MeV plasma with $\lambda = 5.5$. }
\label{fig:D}
\end{figure*}

On the other hand, \cite{Liu:2006ug,D'Eramo:2010ak} computed $\hat{q}_{\text{LRW}} = \pi^{3/2}\Gamma(3/4)\sqrt{\lambda}T^3/\Gamma(5/4)$ from a light-like Wilson loop setup.  Because of the light-like Wilson loop setup, the authors of \cite{Liu:2006ug,D'Eramo:2010ak} claim that $\hat{q}_{\text{LRW}}$ is appropriate for light flavor quark propagation in a strongly-coupled plasma. Notice that $\hat{q}_{\text{LRW}}$ is independent of $\gamma$.

By relating our diffusion coefficient $D(a,d)$ to the transverse momentum fluctuations $\kappa_T$, and subsequently $\hat{q}$, as in \cite{Gubser:2006nz}, we find
\begin{align}
	\label{eq:qhat}
	\hat{q}(a,d) = \frac{32\pi\sqrt{\lambda} T^3}{((d-1)^2(1-\frac{a}{2})v)}.
\end{align}
For $d = 5$ and heavy quarks ($a = 0$), we find the transverse momentum squared imparted to the quark per unit path length from the \emph{thermal medium} is $\hat{q} = 2\pi\sqrt{\lambda}T^3/v$, which reproduces $\hat{q}_{\text{Gubser}}$ without the ``tail wagging the dog'' $\gamma$ dependence originating from the induced worldsheet horizon.  For $d = 5$ and light quarks ($a = 1$) at early times ($v\approx1$), we have $\hat{q} = 4\pi\sqrt{\lambda}T^3$, whose coefficient is numerically similar to, but about a factor of 2/3 larger than, that of $\hat{q}_{\text{LRW}}$.

Our work shows that in AdS/CFT $\qhat$ is independent of quark mass (string length) for times large compared to the inverse temperature of the plasma and is a smooth function of quark velocity $v$.  Thus we gain our first qualitative insight into the motion of heavy quarks above the speed limit: the transverse momentum fluctuations do not change.  Phenomenologically, our work quantitatively demonstrates, for the first time, that AdS/CFT sensibly predicts that the angular correlations of high momentum heavy and light flavor observables at RHIC and LHC should be the same.

\begin{figure*}[!tbp]
\hspace*{\fill}%
\begin{subfigure}[t]{0.48\textwidth}
\centering
\includegraphics[width=\columnwidth]{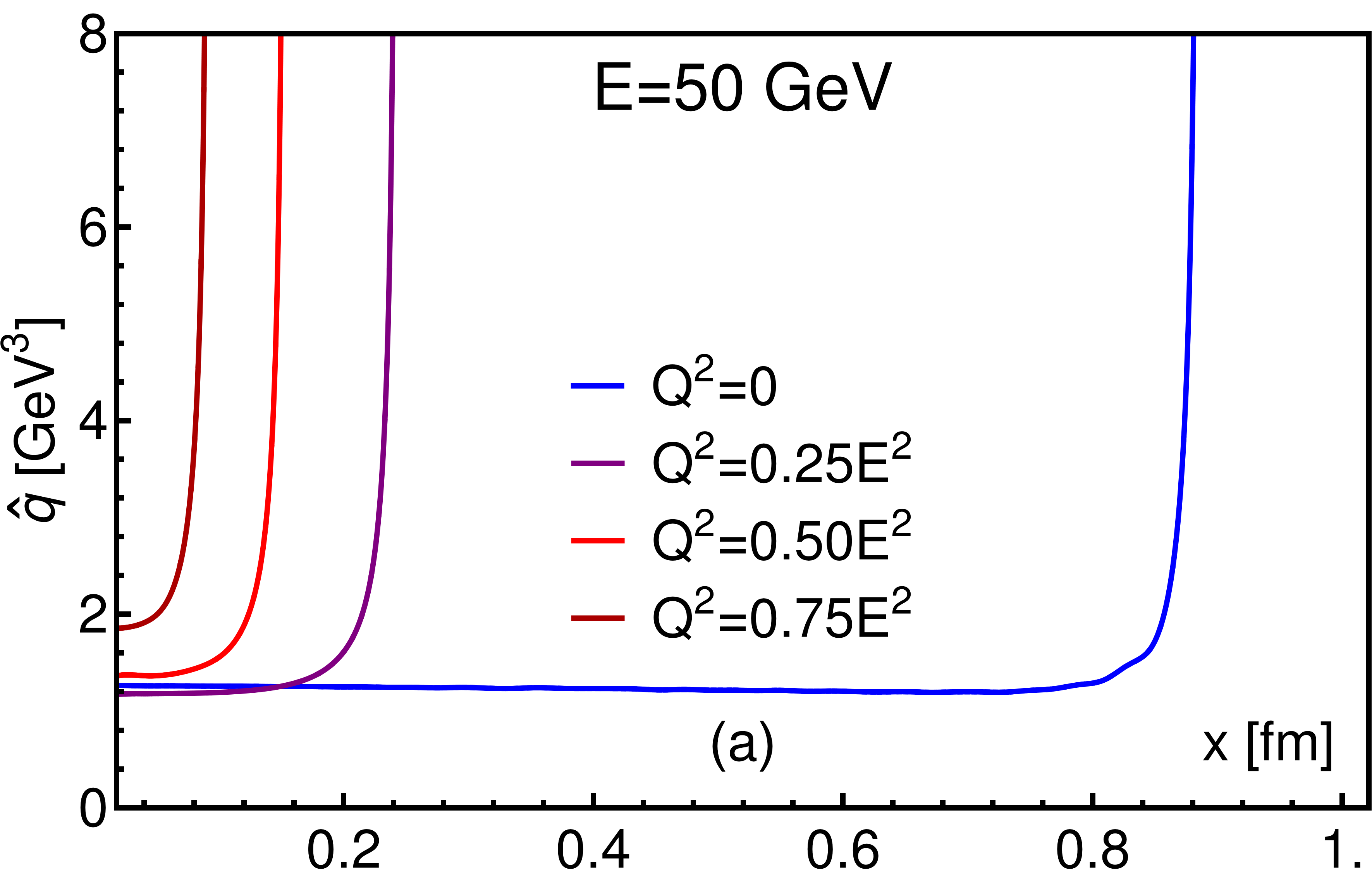}
\end{subfigure}\hspace{.15in}
\begin{subfigure}[t]{0.48\textwidth}
\centering
\includegraphics[width=\columnwidth]{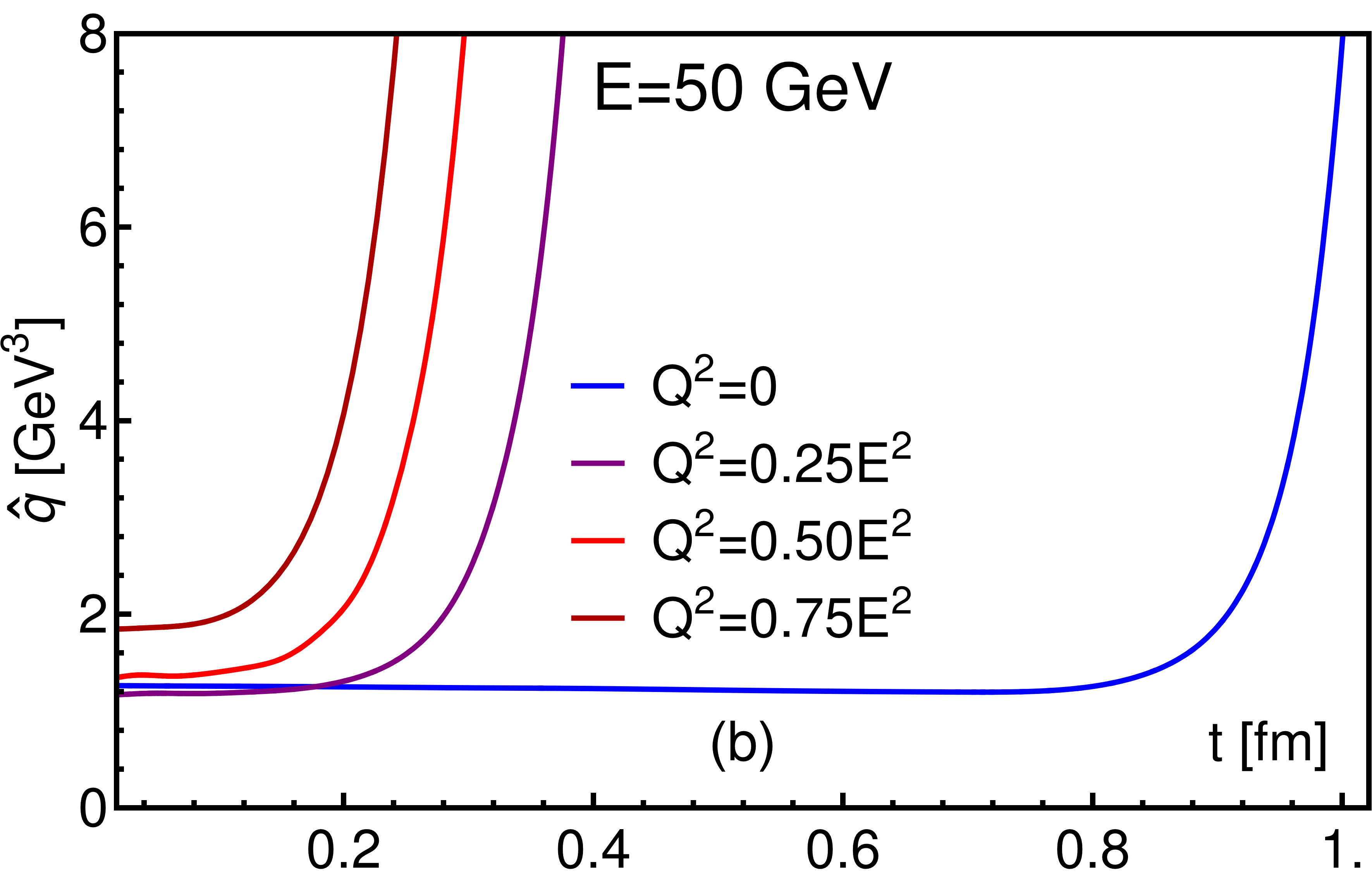}
\end{subfigure}
\hspace*{\fill}\\[.1in]
\hspace*{\fill}%
\begin{subfigure}[t]{0.48\textwidth}
\centering
\includegraphics[width=\columnwidth]{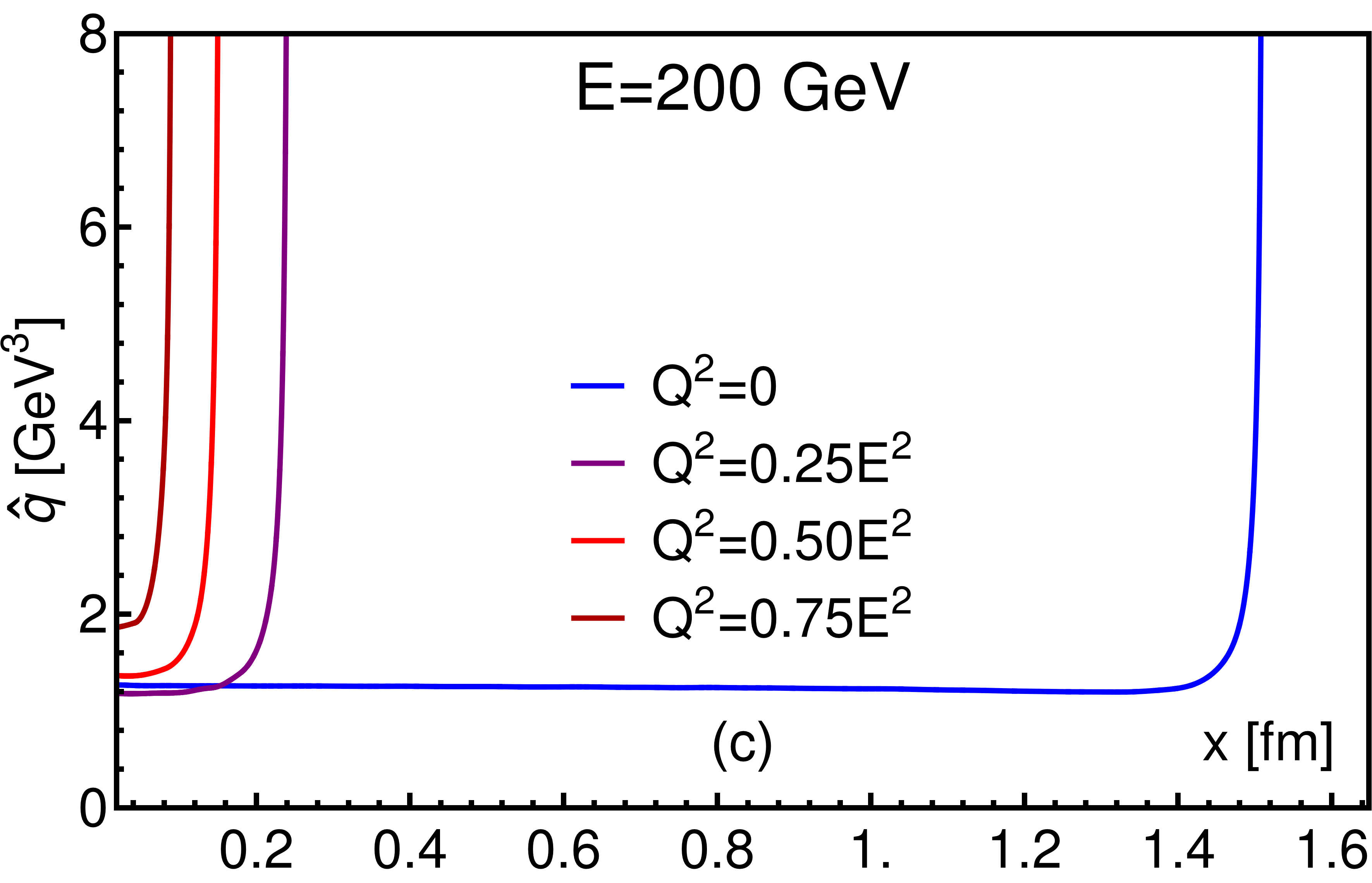}
\end{subfigure}\hspace{.15in}
\begin{subfigure}[t]{0.48\textwidth}
\centering
\includegraphics[width=\columnwidth]{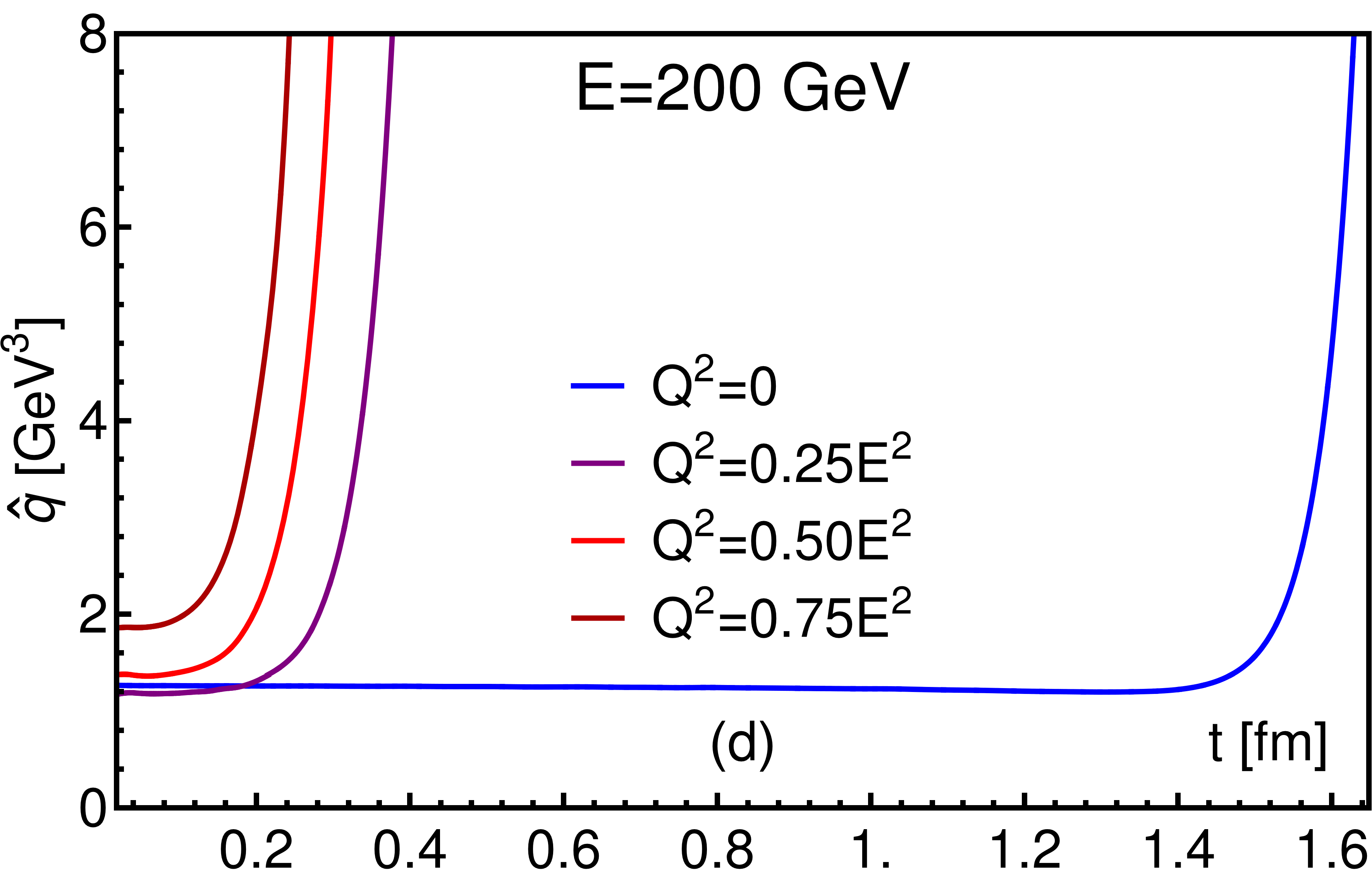}
\end{subfigure}
\hspace*{\fill}
\caption{(Colour online) $\hat{q}$ plotted as a function of $x$ (left column) and $t$ (right column) for an $E =  50$ GeV quark (top row) and $E = 200$ GeV quark (bottom row) for various virtualities in a $T = 350$ MeV plasma with $\lambda = 5.5$.}
\label{fig:qhat}
\end{figure*}

Going further, using the Chesler et al.~\cite{Chesler:2008uy} numerical solutions for a falling string, we may even plot dynamically the $t$ or $x$ dependence of $\hat{q}$ for a light quark in a strongly-coupled plasma.  We show in \reffig{fig:D}, for the first time, the dynamical scaled diffusion coefficient as a function of space, in (a) and (c), and time, in (b) and (d), which one may relate to $v\times\hat{q}$, for quarks of 50 GeV, (a) and (b), and 200 GeV, (c) and (d), of varying virtuality in a $T=350$ MeV plasma typical for RHIC and LHC conditions \cite{Heinz:2013th} and taking $\lambda = 5.5$ \cite{Gubser:2006nz}. The diffusion coefficient smoothly interpolates between that for static heavy quarks ($a = 0$) and light flavor ($a = 1$) as the longitudinal velocity of the string endpoint asymptotes to 0. We emphasize in \reffig{fig:D} an interesting virtuality scaling for the diffusion coefficient with dash-dotted lines, whose analytic origin is under current investigation, that near $t = 0$
\begin{align}
	a(Q^2) \simeq 1 - \frac{1}{2^{4Q^2}}
\end{align}
independent of the original quark energy.

Going beyond the diffusion coefficient, we explicitly plot, for the first time, $\hat{q}$ as a function of position and time in \reffig{fig:qhat}.  At early times, we find $\hat{q}\sim1$ GeV$^3$ independent of initial quark energy, about an order of magnitude larger than estimates from pQCD \cite{Baier:2002tc}, and, necessarily, similar to the original $v\approx1$ AdS/CFT estimate \cite{Liu:2006ug,D'Eramo:2010ak}.  Due to the inverse velocity dependence, the behaviour of $\hat{q}$ at early times becomes non-monotonic as a function of the virtuality.  We see that when the quark begins to slow down $\hat{q}$ increases dramatically, asymptotically approaching infinity when the light flavor energy loss reaches its Bragg peak \cite{Chesler:2008uy,Morad:2014xla}.

One outstanding problem to be solved is to understand the role of the quantum fluctuations of the falling string in the direction of motion of the string endpoints. Recall that in one interpretation, $s^2(t;a)$ corresponded to the average squared transverse distance travelled by an observer travelling down the stretched string at a speed characterised by $a$. One could just as easily consider the average squared transverse distance travelled by an observer travelling down the trailing string of Gubser \cite{Gubser:2006bz} and HKKKY \cite{Herzog:2006gh} at a speed characterised by $a$. This would help in understanding the non-transverse, directional fluctuations of the falling string.

\section*{Acknowledgements}
The authors wish to thank S.\ S.\ Gubser for suggesting the problem as well as J.\ Casalderrey-Solana, D.\ Teaney, J.\ Murugan, J.\ Shock, and I.\ Snyman for insightful discussions. The work of RWM was supported by the Square-Kilometer-Array (SKA) Undergraduate and Honours Bursary Programme. The work of WAH was supported by the National Research Foundation (NRF) and the SA-CERN Collaboration.

\appendix
\section{Static Gauge Polyakov String Equations of Motion\label{app:ply}} 

Recall that in \refsec{sec:lo-poly} we considered the Polyakov action for a bosonic string embedded in $AdS_5$-Schwarzschild 
\begin{align}
S_\text{P} = - \frac{1}{4\pi\alpha^\prime}\int_\manifold d^2\sigma\sqrt{-\gamma}\gamma^{ab}g_{ab} = \int_\manifold d^2\sigma \lagrangian_\text{P}, \label{eq:ply-action}
\end{align}
where $\gamma_{ab}$ is an auxiliary world-sheet metric, $g_{ab} := \partial_aX^\mu\partial_bX^\nu G_{\mu\nu}$ is the induced world-sheet metric, $G_{\mu\nu}$ is the space-time metric for $AdS_5$-Schwarzschild and $X^\mu$ is the embedding of the string world-sheet. Using the associated Polyakov Lagrangian we also defined the canonical momentum conjugate to $X^{\mu}$ by 
\begin{align}
\momenta{a}{\mu} := \frac{\partial\lagrangian_\text{P}}{\partial(\partial_aX^\mu)} = -\frac{1}{2\pi\alpha^\prime}\sqrt{-\gamma}\gamma^{ab}G_{\mu\nu}\partial_bX^\nu. \label{eq:ply-momentum}
\end{align}
Here, we shall provide a more detailed derivation of the string equations of motion obtained in \refeq{eq:lo-ply-eom}. To this end, we require that the functional variation of \refeq{eq:ply-action} with respect to $X^\mu$ vanishes
\begin{align}
\nonumber 0 = (\delta S_\text{P})_{X} &= \frac 12 \int_\manifold d^2\sigma \left((\delta\momenta{a}{\mu})_{X}\partial_aX^\mu + \momenta{a}{\mu}\partial_a(\delta X^\mu)\right) \\
&= \frac 12 \int_\manifold d^2\sigma \left((\delta\momenta{a}{\mu})_{X}\partial_aX^\mu + \partial_a(\momenta{a}{\mu}\delta X^\mu) - \delta X^\mu\partial_a\momenta{a}{\mu}\right). \label{eq:ply-action-var-X1}
\end{align}
The first term in \refeq{eq:ply-action-var-X1} can be simplified as follows:
\begin{align}
\nonumber &(\delta\momenta{a}{\mu})_{X}\partial_aX^\mu = \frac{\partial G_{\mu\nu}}{\partial X^\rho}\delta X^\rho \momenta{a}{\gamma} G^{\gamma\nu}\partial_aX^\mu + \momenta{a}{\mu}\partial_a\delta X^\mu \\
\nonumber &=\frac{\partial G_{\mu\nu}}{\partial X^\rho}\delta X^\rho \momenta{a}{\gamma} G^{\gamma\nu}\partial_aX^\mu + \partial_a(\momenta{a}{\mu}\delta X^\mu) - \delta X^\mu\partial_a\momenta{a}{\mu} \\
&=2\delta X^\rho\Gamma^{\gamma}_{\rho\mu}\momenta{a}{\gamma} \partial_aX^\mu + \partial_a(\momenta{a}{\mu}\delta X^\mu) - \delta X^\mu\partial_a\momenta{a}{\mu},
\label{eq:ply-momentum-var-X}
\end{align}
where in the last line we inserted the Christoffel symbol
\begin{align*}
\Gamma^{\gamma}_{\rho\mu} := \frac 12 G^{\gamma\nu}\left(\partial_\rho G_{\mu\nu} + \partial_\mu G_{\rho\nu} - \partial_\nu G_{\rho\mu}\right).
\end{align*}
Naively, the Christoffel symbol contributes two additional terms in the last line of \refeq{eq:ply-momentum-var-X}, but these additional terms exactly cancel each other since the contraction $\eta^{ab}\partial_aX^\mu\partial_bX^\nu(\partial_\mu G_{\rho\nu} - \partial_\nu G_{\rho\mu})= 0$ vanishes identically. Substituting \refeq{eq:ply-momentum-var-X} into \refeq{eq:ply-action-var-X1} we obtain
\begin{align}
\nonumber 0 = (\delta S_\text{P})_{X} &= -\int_\manifold d^2\sigma\delta X^\mu\left(\partial_a\momenta{a}{\mu} - \Gamma^{\gamma}_{\mu\nu}\momenta{a}{\gamma} \partial_aX^\nu\right) + \int_\manifold d^2\sigma \partial_a(\delta X^\mu\momenta{a}{\mu}) \\
\nonumber &=
-\int_\manifold d^2\sigma\delta X^\mu\left(\partial_a\momenta{a}{\mu} - \Gamma^{\gamma}_{\mu\nu}\momenta{a}{\gamma} \partial_aX^\nu\right) + \int_{\partial\manifold} d\sigma^b \epsilon_{ba}\delta X^\mu\momenta{a}{\mu} \\
&= -\int_\manifold d^2\sigma\delta X^\mu\left(\partial_a\momenta{a}{\mu} - \Gamma^{\gamma}_{\mu\nu}\momenta{a}{\gamma} \partial_aX^\nu\right) + \int_{0}^{t_f} dt\left(\left.\delta X^\mu\momenta{\sigma}{\mu}\right|^{\sigma = \sigma_f}_{\sigma = 0}\right), \label{eq:ply-action-var-X2}
\end{align}
where in going from the second to the third line, we explicitly used $\left.\delta X^\mu\right|_{t\in\{0,t_f\}}=0$.
We can eliminate the boundary integral over $t\in[0,\sigma_f]$ in \refeq{eq:ply-action-var-X2} by choosing appropriate boundary conditions already stipulated in \ref{eq:lo-ply-bcs}, in which case the string equations of motion\footnote{\refeq{eq:ply-eom} is the string analogue of the geodesic equation for point particles. See \cite{carroll2004spacetime}.} are given by 
\begin{align}
0= \partial_a\momenta{a}{\mu} - \Gamma^{\alpha}_{\mu\nu}\partial_a X^\nu\momenta{a}{\alpha} = \nabla_a\momenta{a}{\mu}; \label{eq:ply-eom}
\end{align} 
precisely \refeq{eq:lo-ply-eom}.

\section{Static Gauge Nambu-Goto String Equations of Motion for Transverse Fluctuations\label{app:ng}}

In order to derive in more detail the equations of motion for the transverse fluctuations on top of some LO classical solution, recall
the
effective Polyakov action
\begin{align}
\nonumber S_\text{NG}^{(2)}:&= \frac 12 \int\limits_{\manifold}d^2\sigma \left.\frac{\partial^2 \lagrangian_\text{NG}}{\partial(\partial_aX^I)\partial(\partial_bX^J)}\right|_{X^\mu_0}\partial_aX^I\partial_bX^J \\
&= -\frac{1}{4\pi\alpha^\prime}\int\limits_{\manifold}d^2\sigma\left.\left[\sqrt{-g}g^{ab}G_{IJ}\right]\right|_{X^\mu_0}\partial_aX^I\partial_bX^J = \int\limits_{\manifold}d^2\sigma\lagrangian_\text{NG}^{(2)}, \label{eq:ng-action2}
\end{align} 
where $g_{ab} := \partial_aX^\mu\partial_bX^\nu G_{\mu\nu}$ is the induced world-sheet metric, $G_{\mu\nu}$ is the space-time metric for some $D=(d+1)$-dimensional space-time, $X^\mu_0$ is some leading order classical solution with non-zero components in the $t,r$-plane and $X^I$ are the transverse fluctuations where $I=2,\cdots,d$. The canonical momenta conjugate to the transverse fluctuations was also given by 
\begin{align}
\momenta{a}{I} = \frac{\partial\lagrangian_\text{NG}^{(2)}}{\partial(\partial_aX^I)} = -\frac{1}{2\pi\alpha^\prime}\left.\left[\sqrt{-g}g^{ab}G_{IJ}\right]\right|_{X^\mu_0}\partial_bX^J. \label{eq:ng-momenta}
\end{align}
The equations of motion for the transverse fluctuations are obtained by requiring that the functional variation of \refeq{eq:ng-action2} with respect to $X^I$ vanishes
\begin{align}
\nonumber 0 &= (\delta S_\text{NG}^2)_{X} = -\frac{1}{2\pi\alpha^\prime}\int\limits_{\manifold}d^2\sigma\left.\left[\sqrt{-g}g^{ab}G_{IJ}\right]\right|_{X^\mu_0}\partial_aX^I\partial_b\delta X^J \\
\nonumber &=-\frac{1}{2\pi\alpha^\prime}\int\limits_{\manifold}d^2\sigma\partial_b\left(\left.\left[\sqrt{-g}g^{ab}G_{IJ}\right]\right|_{X^\mu_0}\partial_aX^I\delta X^J\right) \\
\nonumber &\phantom{=} + \frac{1}{2\pi\alpha^\prime}\int\limits_{\manifold}d^2\sigma\partial_b\left(\left.\left[\sqrt{-g}g^{ab}G_{IJ}\right]\right|_{X^\mu_0}\partial_aX^I\right)\delta X^J, \\
\nonumber &=-\frac{1}{2\pi\alpha^\prime}\int\limits_{0}^{t_f}dt\left.\left.\left[\sqrt{-g}g^{ab}G_{IJ}\right]\right|_{X^\mu_0}\partial_aX^I\delta X^J\right|_{\sigma=0}^{\sigma=\sigma_f} \\
&\phantom{=} + \frac{1}{2\pi\alpha^\prime}\int\limits_{\manifold}d^2\sigma\partial_b\left(\left.\left[\sqrt{-g}g^{ab}G_{IJ}\right]\right|_{X^\mu_0}\partial_aX^I\right)\delta X^J,\label{eq:ng-action-var}
\end{align}
where in going from the second to the third line, we explicitly used $\left.\delta X^J\right|_{t\in\{0,t_f\}}=0$. We can eliminate the boundary integral over $t\in[0,t_f]$ in \refeq{eq:ng-action-var} by choosing appropriate boundary conditions stipulate by \ref{eq:nlo-ng-bcs}, in which case the equations of motion for the transverse fluctuations are given by
\begin{align}
0=\partial_b\left(\left.\left[\sqrt{-g}g^{ab}G_{IJ}\right]\right|_{X^\mu_0}\partial_aX^I\right) \label{eq:ng-eom};
\end{align}
precisely \ref{eq:nlo-ng-eom}.

\bibliographystyle{JHEP}
\bibliography{main}
\end{document}